\documentclass[letterpaper,10pt]{article} 
\usepackage{opticameet3} %% use version 3 for proper copyright statement

\newcommand\authormark[1]{\textsuperscript{#1}}

\usepackage{amsmath,amssymb}
\usepackage[colorlinks=true,bookmarks=false,citecolor=blue,urlcolor=blue]{hyperref} %pdflatex
\usepackage{dsfont,color,float,empheq}
\usepackage{graphicx,wrapfig,multicol,multirow}
\usepackage{psfrag,wrapfig}
\usepackage[font=footnotesize]{caption}
\usepackage{pgfplots}
\usepackage{subcaption}
\usepackage{cite}
\usepackage{setspace}
\usepackage{multicol}
\usetikzlibrary{patterns}
\usepackage{tikz,pgfplots}
\usetikzlibrary{positioning,shadows,shapes,backgrounds,calc}
\usepackage{pgfplotstable}

\usepackage[group-separator={,}, mode = match, propagate-math-font = true, reset-math-version = false, reset-text-family = false, reset-text-series = false, text-family-to-math = true, text-series-to-math = true, exponent-product=\cdot, range-phrase=\textellipsis, range-units=single, list-units=single]{siunitx} % SI units. 

\usepackage{csquotes}
\usepackage[UKenglish]{babel}

\usepackage{amsmath}    % Amsmath, helps with a lot of math stuff
\usepackage[protrusion=true,expansion=true]{microtype}
\usepackage{lipsum}
\usepackage{fonttable} % To check the font characters
\usepackage{import}

\usepackage{tikz}
\usepackage{tikz}
\usepackage{fp}
\usepackage{import}
\usetikzlibrary{positioning}
\usetikzlibrary{fit}
\usetikzlibrary{calc}
\usetikzlibrary{shapes}
\usetikzlibrary{math}
\usetikzlibrary{fpu}
\usetikzlibrary{decorations.markings}
\usetikzlibrary{decorations.pathreplacing}
\usetikzlibrary{arrows.meta}
\usetikzlibrary{intersections}

\makeatletter

\def\calcLength(#1,#2)#3{%
\pgfpointdiff{\pgfpointanchor{#1}{center}}%
             {\pgfpointanchor{#2}{center}}%
\pgf@xa=\pgf@x%
\pgf@ya=\pgf@y%
\FPeval\@temp@a{\pgfmath@tonumber{\pgf@xa}}%
\FPeval\@temp@b{\pgfmath@tonumber{\pgf@ya}}%
\FPeval\@temp@sum{(\@temp@a*\@temp@a+\@temp@b*\@temp@b)}%
\FProot{\FPMathLen}{\@temp@sum}{2}%
\FPround\FPMathLen\FPMathLen5\relax
\global\expandafter\edef\csname #3\endcsname{\FPMathLen}
}

\tikzset{outer sep=0}
\tikzset{inner sep=0}

\def\pgfaddtoshape#1#2{%	To add anchors to shape
	\begingroup
	\def\pgf@sm@shape@name{#1}%
	\let\anchor\pgf@sh@anchor
	#2%
	\endgroup
}

\newcommand{\anchorlet}[2]{%	To copy anchors to a different name
	\global\expandafter
	\let\csname pgf@anchor@\pgf@sm@shape@name @#1\expandafter\endcsname
	\csname pgf@anchor@\pgf@sm@shape@name @#2\endcsname
}

\pgfmathsetmacro{\NODESIZE}{42}
\pgfmathsetmacro{\NODETHICKNESS}{1.0}
\pgfmathsetmacro{\ROUNDEDCORNERS}{0.5mm}
\tikzset{node distance=0.5*\NODESIZE pt}

\def\FNODESIZE{\NODESIZE pt}

\tikzstyle{textstyle} = [text height=1.5ex, text depth=.5ex]
\tikzset{every label/.style=textstyle}

\tikzstyle{linestyle} = [line width = \NODETHICKNESS, rounded corners = \ROUNDEDCORNERS]
\tikzstyle{arrowstyle} = [>=stealth, linestyle]
\tikzstyle{<--} = [<-, arrowstyle]
\tikzstyle{-->} = [->, arrowstyle]
\tikzstyle{->-} = [linestyle, decoration={markings,	mark=at position 0.5 with {\arrow[arrowstyle]{>}}}, postaction={decorate}] 
\tikzstyle{-<-} = [linestyle, decoration={markings,	mark=at position 0.5 with {\arrow[arrowstyle]{<}}}, postaction={decorate}] 

\tikzset{   % Modified tikzstyle -A- by Sjoerd
    -A-/.style args={#1}{%
        linestyle, 
        decoration={markings, mark=at position #1 with {\arrow[>=Triangle, scale=.025*\NODESIZE]{>}}},
        postaction={decorate}
    },
    -A-/.default = {0.75}
}

\tikzset{   % Modified tikzstyle -AA- by Menno, draw an actual amplifier (i.e. EDFA style) instead of a solid triangle
    -AA-/.style args={#1, #2, #3}{%
        linestyle,
        decoration={markings, mark=at position #1 with {
            \node[amp={color=#2, width=#3, height=#3}, rotate=\pgfdecoratedangle] at (0,0) (inline_amp) {};
        }},
        postaction={decorate}
    },
    -AA-/.default = {0.5, E, 0.15}
}

\tikzstyle{<-->} = [<->, arrowstyle]
\tikzstyle{---} = [arrowstyle]

\tikzset{
    -PC-/.style args={#1}{
        linestyle, 
        decoration={markings, mark=at position #1 with {
            \draw[---,FO]
                (.05*\NODESIZE*\pgflinewidth,.05*\NODESIZE*\pgflinewidth) circle[radius=.05*\NODESIZE*\pgflinewidth];
            \draw[---,FO]
                (-.05*\NODESIZE*\pgflinewidth,.05*\NODESIZE*\pgflinewidth) circle[radius=.05*\NODESIZE*\pgflinewidth];
            \draw[---,FO]
                (0,-.05*\NODESIZE*\pgflinewidth) circle[radius=.05*\NODESIZE*\pgflinewidth];}},
        postaction={decorate}
        },
        -PC-/.default = {0.5}
}

\definecolor{C0}{RGB}{031, 119, 180} %blue
\definecolor{C1}{RGB}{255, 127, 014} %orange
\definecolor{C2}{RGB}{044, 160, 044} %green
\definecolor{C3}{RGB}{215, 039, 040} %red
\definecolor{C4}{RGB}{148, 103, 189} %purple
\definecolor{C5}{RGB}{140, 086, 075} %brown
\definecolor{C6}{RGB}{227, 119, 194} %pink
\definecolor{C7}{RGB}{127, 127, 127} %grey
\definecolor{C8}{RGB}{188, 189, 034} %olive
\definecolor{C9}{RGB}{023, 190, 207} %cyan

\definecolor{C0l}{RGB}{174, 199, 232} %blue
\definecolor{C1l}{RGB}{255, 187, 120} %orange
\definecolor{C2l}{RGB}{152, 223, 138} %green
\definecolor{C3l}{RGB}{255, 152, 150} %red
\definecolor{C4l}{RGB}{197, 176, 213} %purple
\definecolor{C5l}{RGB}{196, 156, 148} %brown
\definecolor{C6l}{RGB}{247, 182, 210} %pink
\definecolor{C7l}{RGB}{199, 199, 199} %grey
\definecolor{C8l}{RGB}{219, 219, 141} %olive
\definecolor{C9l}{RGB}{158, 218, 229} %cyan

\definecolor{Snow}{HTML}{FBFBFB} 			% Background
\definecolor{TUeRed}{RGB}{200, 25, 25}		% TU/e Red
\definecolor{TUeGreen}{RGB}{25, 200, 113}	% TU/e Green
\definecolor{TUeBlue}{RGB}{25, 113, 200}	% TU/e Blue

\definecolor{O}{RGB}{031, 119, 180} 	% Optical = C0
\definecolor{Ol}{RGB}{174, 199, 232} 	% Optical light
\definecolor{E}{RGB}{255, 127, 014} 	% Electrical = C1
\definecolor{El}{RGB}{255, 187, 120} 	% Electrical light
\definecolor{D}{RGB}{148, 103, 189}     % Digital = C5
\definecolor{Dl}{RGB}{197, 176, 213} 	% Digital light
\definecolor{EO}{RGB}{215, 039, 040} 	% Electro-Optical = C3
\definecolor{EOl}{RGB}{255, 152, 150}   % Electro-Optical light

\tikzstyle{FW} = [fill=white]			% Fill white
\tikzstyle{FB} = [fill=white]			% Fill background, you can override this tikzstyle with the presentation.mplstyle background color (Snow) in your own tikzpicture
\tikzstyle{FO} = [fill=C0l, draw=C0]	% Optical
\tikzstyle{FE} = [fill=C1l, draw=C1]	% Electrical
\tikzstyle{FD} = [fill=C2l, draw=C2]	% Digital

\def\direce{e}
\def\direcw{w}
\def\direcn{n}
\def\direcs{s}

\def\flipfalse{0}
\newcount\portcount
\pgfdeclareshape{basic}{
	\inheritsavedanchors[from=rectangle] 
	
	\inheritanchor[from=rectangle]{center}
	\inheritanchor[from=rectangle]{mid}		
	\inheritanchor[from=rectangle]{base}	
	\inheritanchor[from=rectangle]{north}
	\inheritanchor[from=rectangle]{south}		
	\inheritanchor[from=rectangle]{west}		
	\inheritanchor[from=rectangle]{mid west}				
	\inheritanchor[from=rectangle]{base west}		
	\inheritanchor[from=rectangle]{north west}		
	\inheritanchor[from=rectangle]{south west}		
	\inheritanchor[from=rectangle]{east}
	\inheritanchor[from=rectangle]{mid east}
	\inheritanchor[from=rectangle]{base east}	
	\inheritanchor[from=rectangle]{north east}
	\inheritanchor[from=rectangle]{south east}	
	
	\inheritanchorborder[from=rectangle]	

	\inheritbackgroundpath[from=rectangle]	
}

\pgfaddtoshape{basic}{
	\anchor{west south west}{
		\pgf@process{\northeast}
		\pgf@ya=.5\pgf@y
		\pgf@process{\southwest}
		\pgf@y=1.5\pgf@y
		\advance\pgf@y by \pgf@ya
		\pgf@y=.5\pgf@y
	}
	\anchor{west north west}{
		\pgf@process{\northeast}
		\pgf@ya=1.5\pgf@y
		\pgf@process{\southwest}
		\pgf@y=.5\pgf@y
		\advance\pgf@y by \pgf@ya
		\pgf@y=.5\pgf@y
	}
	\anchor{east north east}{
		\pgf@process{\southwest}
		\pgf@ya=.5\pgf@y
		\pgf@process{\northeast}
		\pgf@y=1.5\pgf@y
		\advance\pgf@y by \pgf@ya
		\pgf@y=.5\pgf@y
	}
	\anchor{east south east}{
		\pgf@process{\southwest}
		\pgf@ya=1.5\pgf@y
		\pgf@process{\northeast}
		\pgf@y=.5\pgf@y
		\advance\pgf@y by \pgf@ya
		\pgf@y=.5\pgf@y
	}
	\anchor{north north west}{
		\pgf@process{\southwest}
		\pgf@xa=1.5\pgf@x
		\pgf@process{\northeast}
		\pgf@x=.5\pgf@x
		\advance\pgf@x by \pgf@xa
		\pgf@x=.5\pgf@x
	}
	\anchor{north north east}{
		\pgf@process{\southwest}
		\pgf@xa=.5\pgf@x
		\pgf@process{\northeast}
		\pgf@x=1.5\pgf@x
		\advance\pgf@x by \pgf@xa
		\pgf@x=.5\pgf@x
	}
	\anchor{south south west}{
		\pgf@process{\northeast}
		\pgf@xa=.5\pgf@x
		\pgf@process{\southwest}
		\pgf@x=1.5\pgf@x
		\advance\pgf@x by \pgf@xa
		\pgf@x=.5\pgf@x
	}
	\anchor{south south east}{
		\pgf@process{\northeast}
		\pgf@xa=1.5\pgf@x
		\pgf@process{\southwest}
		\pgf@x=.5\pgf@x
		\advance\pgf@x by \pgf@xa
		\pgf@x=.5\pgf@x
	}
}

\pgfaddtoshape{basic}{
	\anchorlet{c}{center}		
	\anchorlet{n}{north}
	\anchorlet{e}{east}
	\anchorlet{s}{south}
	\anchorlet{w}{west}			
	\anchorlet{se}{south east}
	\anchorlet{sw}{south west}
	\anchorlet{ne}{north east}
	\anchorlet{nw}{north west}
	\anchorlet{wsw}{west south west}
	\anchorlet{wnw}{west north west}
	\anchorlet{ene}{east north east}
	\anchorlet{ese}{east south east}
	\anchorlet{nnw}{north north west}
	\anchorlet{nne}{north north east}
	\anchorlet{ssw}{south south west}
	\anchorlet{sse}{south south east}
}	% Then the basic shapes

\pgfdeclareshape{ampshape}{
	\savedmacro\direction{
		\edef\direction{\pgfkeysvalueof{/tikz/ampkeys/direction}}%
	}
	\saveddimen\minwidth{
		\pgfmathsetlength\pgf@x{\pgfshapeminwidth}%
	}
	\saveddimen\minheight{
		\pgfmathsetlength\pgf@x{\pgfshapeminheight}%
	}
	\inheritsavedanchors[from=basic] 
	
	\inheritanchor[from=basic]{center}
	\inheritanchor[from=basic]{mid}		
	\inheritanchor[from=basic]{base}	
	\inheritanchor[from=basic]{north}
	\inheritanchor[from=basic]{south}		
	\inheritanchor[from=basic]{west}		
	\inheritanchor[from=basic]{mid west}				
	\inheritanchor[from=basic]{base west}		
	\inheritanchor[from=basic]{north west}		
	\inheritanchor[from=basic]{south west}		
	\inheritanchor[from=basic]{east}
	\inheritanchor[from=basic]{mid east}
	\inheritanchor[from=basic]{base east}	
	\inheritanchor[from=basic]{north east}
	\inheritanchor[from=basic]{south east}
	\inheritanchor[from=basic]{west south west}%
	\inheritanchor[from=basic]{west north west}%
	\inheritanchor[from=basic]{east north east}%
	\inheritanchor[from=basic]{east south east}%
	\inheritanchor[from=basic]{north north west}%
	\inheritanchor[from=basic]{north north east}%
	\inheritanchor[from=basic]{south south west}%
	\inheritanchor[from=basic]{south south east}%		
	
	\inheritanchorborder[from=basic]	
	\inheritbackgroundpath[from=basic]

    \pgfutil@g@addto@macro\pgf@sh@s@ampshape{%
        \pgfutil@ifundefined{pgf@anchor@ampshape@in0}{%	If it is already defined, do not replace
	        \expandafter\xdef\csname pgf@anchor@ampshape@in0\endcsname{%
	            \noexpand\ampshape@port{0}% defined below
	        }%
	    }{}%
        \pgfutil@ifundefined{pgf@anchor@ampshape@in}{%	If it is already defined, do not replace
	        \expandafter\xdef\csname pgf@anchor@ampshape@in\endcsname{%
	            \noexpand\ampshape@port{0}% defined below
	        }%
	    }{}%
        \pgfutil@ifundefined{pgf@anchor@ampshape@out0}{%	If it is already defined, do not replace
	        \expandafter\xdef\csname pgf@anchor@ampshape@out0\endcsname{%
	            \noexpand\ampshape@port{1}% defined below
	        }%
	    }{}%
        \pgfutil@ifundefined{pgf@anchor@ampshape@out}{%	If it is already defined, do not replace
	        \expandafter\xdef\csname pgf@anchor@ampshape@out\endcsname{%
	            \noexpand\ampshape@port{1}% defined below
	        }%
	    }{}%
	}
}

\def\ampshape@port#1{%	#1 defines whether it is input or output
    \northeast	% Everything is defined wrt the northeast anchor

    \ifnum#1=0	% Port on the "left" side, the input
	    \if\direction\direce
			\pgf@x=-\pgf@x
		    \pgf@ya= \pgf@y
		    \pgfmathsetlength{\pgf@y}{\pgf@ya-0.5*\minheight}%
		\fi
	    \if\direction\direcw
			\pgf@x=\pgf@x
		    \pgf@ya= \pgf@y
		    \pgfmathsetlength{\pgf@y}{\pgf@ya-0.5*\minheight}%
		\fi
	    \if\direction\direcn
			\pgf@y=-\pgf@y
		    \pgf@xa=\pgf@x
		    \pgfmathsetlength{\pgf@x}{\pgf@xa-0.5*\minwidth}%
		\fi
	    \if\direction\direcs
			\pgf@y=\pgf@y
		    \pgf@xa= \pgf@x
		    \pgfmathsetlength{\pgf@x}{\pgf@xa-0.5*\minwidth}%
		\fi
	\else	% "Right" port, the output
	    \if\direction\direce
			\pgf@x=\pgf@x
		    \pgf@ya= \pgf@y
		    \pgfmathsetlength{\pgf@y}{\pgf@ya-0.5*\minheight}%
		\fi
	    \if\direction\direcw
			\pgf@x=-\pgf@x
		    \pgf@ya= \pgf@y
		    \pgfmathsetlength{\pgf@y}{\pgf@ya-0.5*\minheight}%
		\fi
	    \if\direction\direcn
			\pgf@y=\pgf@y
		    \pgf@xa= \pgf@x
		    \pgfmathsetlength{\pgf@x}{\pgf@xa-0.5*\minwidth}%
		\fi
	    \if\direction\direcs
			\pgf@y=-\pgf@y
		    \pgf@xa= \pgf@x
		    \pgfmathsetlength{\pgf@x}{\pgf@xa-0.5*\minwidth}%
		\fi
	\fi
}

\pgfaddtoshape{ampshape}{
	\anchorlet{c}{center}		
	\anchorlet{n}{north}
	\anchorlet{e}{east}
	\anchorlet{s}{south}
	\anchorlet{w}{west}			
	\anchorlet{se}{south east}
	\anchorlet{sw}{south west}
	\anchorlet{ne}{north east}
	\anchorlet{nw}{north west}
	\anchorlet{wsw}{west south west}
	\anchorlet{wnw}{west north west}
	\anchorlet{ene}{east north east}
	\anchorlet{ese}{east south east}
	\anchorlet{nnw}{north north west}
	\anchorlet{nne}{north north east}
	\anchorlet{ssw}{south south west}
	\anchorlet{sse}{south south east}
}

\tikzset{
	/tikz/ampkeys/.cd,
	height/.initial=0.5,
	width/.initial=0.5,
	color/.initial=O,
	direction/.initial=e,
	linestyle/.initial={linestyle, rounded corners = 0},
	/tikz/amp/.code={
		\pgfqkeys{/tikz/ampkeys}{#1}%
		\tikzset{/tikz/ampkeys/drawer/.expanded=%
			{\pgfkeysvalueof{/tikz/ampkeys/direction}}%
			{\pgfkeysvalueof{/tikz/ampkeys/height}}%
			{\pgfkeysvalueof{/tikz/ampkeys/width}}%
			{\pgfkeysvalueof{/tikz/ampkeys/color}}%
			{\pgfkeysvalueof{/tikz/ampkeys/linestyle}}%
		}
	},
	/tikz/ampkeys/drawer/.code n args={5}{%
		\tikzset{
			ampshape,
			minimum height=#2*\NODESIZE,
			minimum width=#3*\NODESIZE,
			append after command={
				\pgfextra{\let\bdr=\tikzlastnode%
				\if#1e
					\draw[draw=#4, fill=#4l, #5] (\bdr.sw) to (\bdr.nw) to (\bdr.e) to cycle {};
				\fi
				\if#1w
					\draw[draw=#4, fill=#4l, #5] (\bdr.se) to (\bdr.ne) to (\bdr.w) to cycle {};
				\fi
				\if#1n
					\draw[draw=#4, fill=#4l, #5] (\bdr.se) to (\bdr.sw) to (\bdr.n) to cycle {};
				\fi
				\if#1s
					\draw[draw=#4, fill=#4l, #5] (\bdr.ne) to (\bdr.nw) to (\bdr.s) to cycle {};
				\fi
				}
			}
		}
	},
}
\newcount\portcount

\pgfdeclareshape{aomshape}{
	\savedmacro\direction{
		\edef\direction{\pgfkeysvalueof{/tikz/aomkeys/direction}}%
	}
	\saveddimen\minwidth{
		\pgfmathsetlength\pgf@x{\pgfshapeminwidth}%
	}
	\saveddimen\minheight{
		\pgfmathsetlength\pgf@x{\pgfshapeminheight}%
	}
	\inheritsavedanchors[from=basic]

	\inheritanchor[from=basic]{center}
	\inheritanchor[from=basic]{mid}
	\inheritanchor[from=basic]{base}
	\inheritanchor[from=basic]{north}
	\inheritanchor[from=basic]{south}
	\inheritanchor[from=basic]{west}
	\inheritanchor[from=basic]{mid west}
	\inheritanchor[from=basic]{base west}
	\inheritanchor[from=basic]{north west}
	\inheritanchor[from=basic]{south west}
	\inheritanchor[from=basic]{east}
	\inheritanchor[from=basic]{mid east}
	\inheritanchor[from=basic]{base east}
	\inheritanchor[from=basic]{north east}
	\inheritanchor[from=basic]{south east}
	\inheritanchor[from=basic]{west south west}%
	\inheritanchor[from=basic]{west north west}%
	\inheritanchor[from=basic]{east north east}%
	\inheritanchor[from=basic]{east south east}%
	\inheritanchor[from=basic]{north north west}%
	\inheritanchor[from=basic]{north north east}%
	\inheritanchor[from=basic]{south south west}%
	\inheritanchor[from=basic]{south south east}%		

	\inheritanchorborder[from=basic]
	\inheritbackgroundpath[from=basic]

	\pgfutil@g@addto@macro\pgf@sh@s@aomshape{%
		\pgfutil@ifundefined{pgf@anchor@aomshape@in0}{%	If it is already defined, do not replace
			\expandafter\xdef\csname pgf@anchor@aomshape@in0\endcsname{%
				\noexpand\aomshape@port{0}% defined below
			}%
		}{}%
		\pgfutil@ifundefined{pgf@anchor@aomshape@in}{%	If it is already defined, do not replace
			\expandafter\xdef\csname pgf@anchor@aomshape@in\endcsname{%
				\noexpand\aomshape@port{0}% defined below
			}%
		}{}%
		\pgfutil@ifundefined{pgf@anchor@aomshape@out0}{%	If it is already defined, do not replace
			\expandafter\xdef\csname pgf@anchor@aomshape@out0\endcsname{%
				\noexpand\aomshape@port{1}% defined below
			}%
		}{}%
		\pgfutil@ifundefined{pgf@anchor@aomshape@out}{%	If it is already defined, do not replace
			\expandafter\xdef\csname pgf@anchor@aomshape@out\endcsname{%
				\noexpand\aomshape@port{1}% defined below
			}%
		}{}%
	}
}

\def\aomshape@port#1{%	#1 defines whether it is input or output
	\northeast	% Everything is defined wrt the northeast anchor

	\ifnum#1=0	% Port on the "left" side, the input
		\if\direction\direce
			\pgf@x=-\pgf@x
			\pgf@ya= \pgf@y
			\pgfmathsetlength{\pgf@y}{\pgf@ya-0.5*\minheight}%
		\fi
		\if\direction\direcw
			\pgf@x=\pgf@x
			\pgf@ya= \pgf@y
			\pgfmathsetlength{\pgf@y}{\pgf@ya-0.5*\minheight}%
		\fi
		\if\direction\direcn
			\pgf@y=-\pgf@y
			\pgf@xa=\pgf@x
			\pgfmathsetlength{\pgf@x}{\pgf@xa-0.5*\minwidth}%
		\fi
		\if\direction\direcs
			\pgf@y=\pgf@y
			\pgf@xa= \pgf@x
			\pgfmathsetlength{\pgf@x}{\pgf@xa-0.5*\minwidth}%
		\fi
	\else	% "Right" port, the output
		\if\direction\direce
			\pgf@x=\pgf@x
			\pgf@ya= \pgf@y
			\pgfmathsetlength{\pgf@y}{\pgf@ya-0.5*\minheight}%
		\fi
		\if\direction\direcw
			\pgf@x=-\pgf@x
			\pgf@ya= \pgf@y
			\pgfmathsetlength{\pgf@y}{\pgf@ya-0.5*\minheight}%
		\fi
		\if\direction\direcn
			\pgf@y=\pgf@y
			\pgf@xa= \pgf@x
			\pgfmathsetlength{\pgf@x}{\pgf@xa-0.5*\minwidth}%
		\fi
		\if\direction\direcs
			\pgf@y=-\pgf@y
			\pgf@xa= \pgf@x
			\pgfmathsetlength{\pgf@x}{\pgf@xa-0.5*\minwidth}%
		\fi
	\fi
}

\pgfaddtoshape{aomshape}{
	\anchorlet{c}{center}
	\anchorlet{n}{north}
	\anchorlet{e}{east}
	\anchorlet{s}{south}
	\anchorlet{w}{west}
	\anchorlet{se}{south east}
	\anchorlet{sw}{south west}
	\anchorlet{ne}{north east}
	\anchorlet{nw}{north west}
	\anchorlet{wsw}{west south west}
	\anchorlet{wnw}{west north west}
	\anchorlet{ene}{east north east}
	\anchorlet{ese}{east south east}
	\anchorlet{nnw}{north north west}
	\anchorlet{nne}{north north east}
	\anchorlet{ssw}{south south west}
	\anchorlet{sse}{south south east}
}

\tikzset{
/tikz/aomkeys/.cd,
size/.initial=1,
circlesize/.initial=1,
color/.initial=O,
direction/.initial=e,
linestyle/.initial={linestyle, inner sep=0.5mm},
fillgradient/.initial=O,
/tikz/aom/.code={
\pgfqkeys{/tikz/aomkeys}{#1}%
\tikzset{/tikz/aomkeys/drawer/.expanded=%
	{\pgfkeysvalueof{/tikz/aomkeys/size}}%
	{\pgfkeysvalueof{/tikz/aomkeys/color}}%
	{\pgfkeysvalueof{/tikz/aomkeys/linestyle}}%
\if\pgfkeysvalueof{/tikz/aomkeys/direction}e
	{0}%
\fi
\if\pgfkeysvalueof{/tikz/aomkeys/direction}w
	{0}%
\fi
\if\pgfkeysvalueof{/tikz/aomkeys/direction}n
	{1}%
\fi
\if\pgfkeysvalueof{/tikz/aomkeys/direction}s
	{1}%
\fi
{\pgfkeysvalueof{/tikz/aomkeys/direction}}%
{\pgfkeysvalueof{/tikz/aomkeys/circlesize}}%
{\pgfkeysvalueof{/tikz/aomkeys/fillgradient}}%
}
},
/tikz/aomkeys/drawer/.code n args={7}{%
		\tikzset{
			aomshape,
			draw,
			minimum height = #1*\NODESIZE,
			minimum width = #1*\NODESIZE,
			#2,
			#3,
			append after command={
					\pgfextra{\let\bdr=\tikzlastnode%
						\node[#7, fit=(\bdr.nw)(\bdr.se)] (boxgradient){};

						\node[coordinate] at ($(\bdr.in)!0.25!(\bdr.out)$) (circlein){};
						\node[coordinate] at ($(\bdr.in)!0.75!(\bdr.out)$) (circleout){};

						\ifnum#4>0
							\node[coordinate] at (circleout -| \bdr.nne) (circleouttop){};
						\else
							\node[coordinate] at (circleout |- \bdr.ene) (circleouttop){};
						\fi

						\draw[---, #2, #3, fill] (\bdr.in) to (circlein) circle (0.05*#6);
						\draw[---, #2, #3, fill] (\bdr.out) to (circleout) circle (0.05*#6);

						\draw[---, #2, #3] (circlein) to (circleouttop){};
					}
				}
		}
	},
}
\newcount\portcount

\pgfdeclareshape{boxshape}{
	\savedmacro\nin{
		\edef\nin{\pgfkeysvalueof{/tikz/boxkeys/nin}}%
	}
	\savedmacro\nout{
		\edef\nout{\pgfkeysvalueof{/tikz/boxkeys/nout}}%
	}
	\savedmacro\direction{
		\edef\direction{\pgfkeysvalueof{/tikz/boxkeys/direction}}%
	}
	\inheritsavedanchors[from=basic]

	\inheritanchor[from=basic]{center}
	\inheritanchor[from=basic]{mid}
	\inheritanchor[from=basic]{base}
	\inheritanchor[from=basic]{north}
	\inheritanchor[from=basic]{south}
	\inheritanchor[from=basic]{west}
	\inheritanchor[from=basic]{mid west}
	\inheritanchor[from=basic]{base west}
	\inheritanchor[from=basic]{north west}
	\inheritanchor[from=basic]{south west}
	\inheritanchor[from=basic]{east}
	\inheritanchor[from=basic]{mid east}
	\inheritanchor[from=basic]{base east}
	\inheritanchor[from=basic]{north east}
	\inheritanchor[from=basic]{south east}
	\inheritanchor[from=basic]{west south west}%
	\inheritanchor[from=basic]{west north west}%
	\inheritanchor[from=basic]{east north east}%
	\inheritanchor[from=basic]{east south east}%
	\inheritanchor[from=basic]{north north west}%
	\inheritanchor[from=basic]{north north east}%
	\inheritanchor[from=basic]{south south west}%
	\inheritanchor[from=basic]{south south east}%		

	\inheritanchorborder[from=basic]
	\inheritbackgroundpath[from=basic]

	\pgfutil@g@addto@macro\pgf@sh@s@boxshape{%
		\pgfmathsetcount{\portcount}{0}
		\pgfmathloop%
		\ifnum\the\portcount<\nin%	Loop over the amount of ports
		\pgfutil@ifundefined{pgf@anchor@boxshape@in\the\portcount}{%	If it is already defined, do not replace
			\expandafter\xdef\csname pgf@anchor@boxshape@in\the\portcount\endcsname{%
				\noexpand\boxshape@port[\the\portcount]{0}% defined below
			}%
		}{}%
		\ifnum\the\portcount=0%	Make a general in anchor 
			\pgfutil@ifundefined{pgf@anchor@boxshape@in}{%
				\expandafter\xdef\csname pgf@anchor@boxshape@in\endcsname{%
					\noexpand\boxshape@port[\the\portcount]{0}% defined below
				}%
			}{}%
		\fi
		\pgfmathaddtocount{\portcount}{1}	% Portcount += 1
		\repeatpgfmathloop%					% GOTO \pgfmathloop
		\pgfmathsetcount{\portcount}{0}
		\pgfmathloop%	        
		\ifnum\the\portcount<\nout% Add output ports, out0, out1, out2, ... up to \nout
		\pgfutil@ifundefined{pgf@anchor@boxshape@out\the\portcount}{%
			\expandafter\xdef\csname pgf@anchor@boxshape@out\the\portcount\endcsname{%
				\noexpand\boxshape@port[\the\portcount]{1}% defined below
			}%
		}{}%
		\ifnum\the\portcount=0%	Make a general out anchor 
			\pgfutil@ifundefined{pgf@anchor@boxshape@out}{%
				\expandafter\xdef\csname pgf@anchor@boxshape@out\endcsname{%
					\noexpand\boxshape@port[\the\portcount]{1}% defined below
				}%
			}{}%
		\fi
		\pgfmathaddtocount{\portcount}{1}	% Portcount += 1
		\repeatpgfmathloop%					% GOTO \pgfmathloop
	}
}

\def\boxshape@port[#1]#2{%	#1 is the port number, #2 defines whether it is input or output
	\northeast \pgf@xa=\pgf@x \pgf@ya=\pgf@y% Store northeast in xa and ya
	\southwest \pgf@xb=\pgf@x \pgf@yb=\pgf@y% Store southwest in xb and yb

	\ifnum#2=0	% Input ports
		\if\direction\direce	% Direction E is the normal direction since most setup diagrams go from left to right
			\pgf@x=\pgf@xb
			\pgf@yc=\pgf@ya \advance\pgf@yc by -\pgf@yb	% yc = ya - yb
			\pgfmathsetlength{\pgf@y}{\pgf@ya-(#1 + 0.5)*(\pgf@yc/\nin)}%
		\fi
		\if\direction\direcw
			\pgf@x=\pgf@xa
			\pgf@yc=\pgf@ya \advance\pgf@yc by -\pgf@yb	% yc = ya - yb
			\pgfmathsetlength{\pgf@y}{\pgf@ya-(#1 + 0.5)*(\pgf@yc/\nin)}%
		\fi
		\if\direction\direcn
			\pgf@y=\pgf@yb
			\pgf@xc=\pgf@xa \advance\pgf@xc by -\pgf@xb	% xc = xa - xb
			\pgfmathsetlength{\pgf@x}{\pgf@xb+(#1 + 0.5)*(\pgf@xc/\nin)}%
		\fi
		\if\direction\direcs
			\pgf@y=\pgf@ya
			\pgf@xc=\pgf@xa \advance\pgf@xc by -\pgf@xb	% xc = xa - xb
			\pgfmathsetlength{\pgf@x}{\pgf@xb+(#1 + 0.5)*(\pgf@xc/\nin)}%
		\fi
	\else	% Output ports
		\if\direction\direce	% Direction E is the normal direction since most setup diagrams go from left to right
			\pgf@x=\pgf@xa
			\pgf@yc=\pgf@ya \advance\pgf@yc by -\pgf@yb	% yc = ya - yb
			\pgfmathsetlength{\pgf@y}{\pgf@ya-(#1 + 0.5)*(\pgf@yc/\nout)}%
		\fi
		\if\direction\direcw
			\pgf@x=\pgf@xb
			\pgf@yc=\pgf@ya \advance\pgf@yc by -\pgf@yb	% yc = ya - yb
			\pgfmathsetlength{\pgf@y}{\pgf@ya-(#1 + 0.5)*(\pgf@yc/\nout)}%
		\fi
		\if\direction\direcn
			\pgf@y=\pgf@ya
			\pgf@xc=\pgf@xa \advance\pgf@xc by -\pgf@xb	% xc = xa - xb
			\pgfmathsetlength{\pgf@x}{\pgf@xb+(#1 + 0.5)*(\pgf@xc/\nout)}%
		\fi
		\if\direction\direcs
			\pgf@y=\pgf@yb
			\pgf@xc=\pgf@xa \advance\pgf@xc by -\pgf@xb	% xc = xa - xb
			\pgfmathsetlength{\pgf@x}{\pgf@xb+(#1 + 0.5)*(\pgf@xc/\nout)}%
		\fi
	\fi
}

\pgfaddtoshape{boxshape}{
	\anchorlet{c}{center}
	\anchorlet{n}{north}
	\anchorlet{e}{east}
	\anchorlet{s}{south}
	\anchorlet{w}{west}
	\anchorlet{se}{south east}
	\anchorlet{sw}{south west}
	\anchorlet{ne}{north east}
	\anchorlet{nw}{north west}
	\anchorlet{wsw}{west south west}
	\anchorlet{wnw}{west north west}
	\anchorlet{ene}{east north east}
	\anchorlet{ese}{east south east}
	\anchorlet{nnw}{north north west}
	\anchorlet{nne}{north north east}
	\anchorlet{ssw}{south south west}
	\anchorlet{sse}{south south east}
}

\tikzset{
/tikz/boxkeys/.cd,
height/.initial=0.5,
width/.initial=1,
color/.initial=O,
direction/.initial=e,
linestyle/.initial={linestyle, inner sep=0.5mm},
nin/.initial=1,
nout/.initial=1,
draw/.initial=1,
/tikz/box/.code={
\pgfqkeys{/tikz/boxkeys}{#1}%
\tikzset{/tikz/boxkeys/drawer/.expanded=%
\if\pgfkeysvalueof{/tikz/boxkeys/direction}e
	{\pgfkeysvalueof{/tikz/boxkeys/width}}%
	{\pgfkeysvalueof{/tikz/boxkeys/height}}%
	{0}%
	{-90}%
\fi
\if\pgfkeysvalueof{/tikz/boxkeys/direction}w
	{\pgfkeysvalueof{/tikz/boxkeys/width}}%
	{\pgfkeysvalueof{/tikz/boxkeys/height}}%
	{0}%
	{90}%
\fi
\if\pgfkeysvalueof{/tikz/boxkeys/direction}n
	{\pgfkeysvalueof{/tikz/boxkeys/width}}%
	{\pgfkeysvalueof{/tikz/boxkeys/height}}%
	{1}%
	{0}%
\fi
\if\pgfkeysvalueof{/tikz/boxkeys/direction}s
	{\pgfkeysvalueof{/tikz/boxkeys/width}}%
	{\pgfkeysvalueof{/tikz/boxkeys/height}}%
	{1}%
	{180}%
\fi
{\pgfkeysvalueof{/tikz/boxkeys/color}}%
{\pgfkeysvalueof{/tikz/boxkeys/linestyle}}%
\ifnum\pgfkeysvalueof{/tikz/boxkeys/draw}>0%
	{draw}%
\else
	{}
\fi
}
},
/tikz/boxkeys/drawer/.code n args={7}{%
		\tikzset{
			boxshape,
			#7,
			#6,
			#5,
			minimum height=
			\ifnum#3>0	% Flip width and height if #3 is not 0
				#1*\NODESIZE
			\else
				#2*\NODESIZE
			\fi
			,minimum width=
			\ifnum#3>0
				#2*\NODESIZE
			\else
				#1*\NODESIZE
			\fi
		}
	},
}
\pgfdeclareshape{beshape}{ % Beam expander shape
    \savedmacro\nports{
        \edef\nports{\pgfkeysvalueof{/tikz/bekeys/nports}}%
    }
    \savedmacro\direction{
        \edef\direction{\pgfkeysvalueof{/tikz/bekeys/direction}}%
    }
    \savedmacro\inverted{
        \edef\inverted{\pgfkeysvalueof{/tikz/bekeys/inverted}}%
    }
    \savedmacro\ninports{
        \ifnum\inverted=0
        \edef\ninports{\nports}
        \else
        \edef\ninports{1}%
        \fi
    }
    \savedmacro\noutports{
        \ifnum\inverted=0
        \edef\noutports{1}%
        \else
        \edef\noutports{\nports}%
        \fi
    }
    \inheritsavedanchors[from=basic]

    \inheritanchor[from=basic]{center}
    \inheritanchor[from=basic]{mid}
    \inheritanchor[from=basic]{base}
    \inheritanchor[from=basic]{north}
    \inheritanchor[from=basic]{south}
    \inheritanchor[from=basic]{west}
    \inheritanchor[from=basic]{mid west}
    \inheritanchor[from=basic]{base west}
    \inheritanchor[from=basic]{north west}
    \inheritanchor[from=basic]{south west}
    \inheritanchor[from=basic]{east}
    \inheritanchor[from=basic]{mid east}
    \inheritanchor[from=basic]{base east}
    \inheritanchor[from=basic]{north east}
    \inheritanchor[from=basic]{south east}
    \inheritanchor[from=basic]{west south west}%
    \inheritanchor[from=basic]{west north west}%
    \inheritanchor[from=basic]{east north east}%
    \inheritanchor[from=basic]{east south east}%
    \inheritanchor[from=basic]{north north west}%
    \inheritanchor[from=basic]{north north east}%
    \inheritanchor[from=basic]{south south west}%
    \inheritanchor[from=basic]{south south east}%

    \inheritanchorborder[from=basic]
    \inheritbackgroundpath[from=basic]

    \pgfutil@g@addto@macro\pgf@sh@s@beshape{%
        \pgfmathsetcount{\portcount}{0}
        \pgfmathloop%
        \ifnum\the\portcount<\nports%	Loop over the amount of ports
        \ifnum\the\portcount<\ninports%	Add input ports, in0, in1, in2, ... up to \ninports
        \pgfutil@ifundefined{pgf@anchor@beshape@in\the\portcount}{%	If it is already defined, do not replace
            \expandafter\xdef\csname pgf@anchor@beshape@in\the\portcount\endcsname{%
                \noexpand\beshape@port[\the\portcount]{0}% defined below
            }%
        }{}%
        \ifnum\the\portcount=0%	Make a general in anchor
        \pgfutil@ifundefined{pgf@anchor@beshape@in}{%
            \expandafter\xdef\csname pgf@anchor@beshape@in\endcsname{%
                \noexpand\beshape@port[\the\portcount]{0}% defined below
            }%
        }{}%
        \fi
        \fi
        \ifnum\the\portcount<\noutports% Add output ports, out0, out1, out2, ... up to \noutports
        \pgfutil@ifundefined{pgf@anchor@beshape@out\the\portcount}{%
            \expandafter\xdef\csname pgf@anchor@beshape@out\the\portcount\endcsname{%
                \noexpand\beshape@port[\the\portcount]{1}% defined below
            }%
        }{}%
        \ifnum\the\portcount=0%	Make a general out anchor
        \pgfutil@ifundefined{pgf@anchor@beshape@out}{%
            \expandafter\xdef\csname pgf@anchor@beshape@out\endcsname{%
                \noexpand\beshape@port[\the\portcount]{1}% defined below
            }%
        }{}%
        \fi
        \fi
        \pgfmathaddtocount{\portcount}{1}    % Portcount += 1
        \repeatpgfmathloop%					% GOTO \pgfmathloop
    }
}

\def\beshape@port[#1]#2{%	#1 is the port number, #2 defines whether it is input or output
    \northeast \pgf@xa=\pgf@x \pgf@ya=\pgf@y% Store northeast in xa and ya
    \southwest \pgf@xb=\pgf@x \pgf@yb=\pgf@y% Store southwest in xb and yb

    \ifnum#2=0
    \ifnum\inverted=0
    \def\chooseports{0}
    \else
    \def\chooseports{1}
    \fi
    \else
    \ifnum\inverted=0
    \def\chooseports{1}
    \else
    \def\chooseports{0}
    \fi
    \fi

    \ifnum\chooseports=0    % Ports on the "left" side, usually the inputs
    \if\direction\direce
    \pgf@x=\pgf@xb
    \pgf@yc=\pgf@ya \advance\pgf@yc by -\pgf@yb    % yc = ya - yb
    \pgfmathsetlength{\pgf@y}{\pgf@ya-(#1 + 0.5)*(\pgf@yc/\nports)}%
    \fi
    \if\direction\direcw
    \pgf@x=\pgf@xa
    \pgf@yc=\pgf@ya \advance\pgf@yc by -\pgf@yb    % yc = ya - yb
    \pgfmathsetlength{\pgf@y}{\pgf@ya-(#1 + 0.5)*(\pgf@yc/\nports)}%
    \fi
    \if\direction\direcn
    \pgf@y=\pgf@yb
    \pgf@xc=\pgf@xa \advance\pgf@xc by -\pgf@xb    % xc = xa - xb
    \pgfmathsetlength{\pgf@x}{\pgf@xb+(#1 + 0.5)*(\pgf@xc/\nports)}%
    \fi
    \if\direction\direcs
    \pgf@y=\pgf@ya
    \pgf@xc=\pgf@xa \advance\pgf@xc by -\pgf@xb    % xc = xa - xb
    \pgfmathsetlength{\pgf@x}{\pgf@xb+(#1 + 0.5)*(\pgf@xc/\nports)}%
    \fi
    \else    % Common port, usually the outputs
    \if\direction\direce
    \pgf@x=\pgf@xa
    \pgf@yc=\pgf@ya \advance\pgf@yc by -\pgf@yb    % yc = ya - yb
    \pgfmathsetlength{\pgf@y}{\pgf@ya-0.5\pgf@yc}%
    \fi
    \if\direction\direcw
    \pgf@x=\pgf@xb
    \pgf@yc=\pgf@ya \advance\pgf@yc by -\pgf@yb    % yc = ya - yb
    \pgfmathsetlength{\pgf@y}{\pgf@ya-0.5\pgf@yc}%
    \fi
    \if\direction\direcn
    \pgf@y=\pgf@ya
    \pgf@xc=\pgf@xa \advance\pgf@xc by -\pgf@xb    % xc = xa - xb
    \pgfmathsetlength{\pgf@x}{\pgf@xa-0.5\pgf@xc}%
    \fi
    \if\direction\direcs
    \pgf@y=\pgf@yb
    \pgf@xc=\pgf@xa \advance\pgf@xc by -\pgf@xb    % xc = xa - xb
    \pgfmathsetlength{\pgf@x}{\pgf@xa-0.5\pgf@xc}%
    \fi
    \fi
}

\pgfaddtoshape{beshape}{
    \anchorlet{c}{center}
    \anchorlet{n}{north}
    \anchorlet{e}{east}
    \anchorlet{s}{south}
    \anchorlet{w}{west}
    \anchorlet{se}{south east}
    \anchorlet{sw}{south west}
    \anchorlet{ne}{north east}
    \anchorlet{nw}{north west}
    \anchorlet{wsw}{west south west}
    \anchorlet{wnw}{west north west}
    \anchorlet{ene}{east north east}
    \anchorlet{ese}{east south east}
    \anchorlet{nnw}{north north west}
    \anchorlet{nne}{north north east}
    \anchorlet{ssw}{south south west}
    \anchorlet{sse}{south south east}
}

\tikzset{
    /tikz/bekeys/.cd,
    height/.initial=1,
    width/.initial=0.5,
    color/.initial=O,
    direction/.initial=e,
    linestyle/.initial={linestyle, rounded corners = 0},
    nports/.initial=3,
    inverted/.initial=0,    % The invert key can be used to make a debe
    sbe/.initial=0,
    angle/.initial=60,
    fillgradient/.initial=O,
    /tikz/be/.code={
        \pgfqkeys{/tikz/bekeys}{#1}%
        \tikzset{/tikz/bekeys/drawer/.expanded=%
            \if\pgfkeysvalueof{/tikz/bekeys/direction}e
                {\pgfkeysvalueof{/tikz/bekeys/width}}%
                {\pgfkeysvalueof{/tikz/bekeys/height}}%
                {0}%
                {-90}%
            \fi
            \if\pgfkeysvalueof{/tikz/bekeys/direction}w
                {\pgfkeysvalueof{/tikz/bekeys/width}}%
                {\pgfkeysvalueof{/tikz/bekeys/height}}%
                {0}%
                {90}%
            \fi
            \if\pgfkeysvalueof{/tikz/bekeys/direction}n
                {\pgfkeysvalueof{/tikz/bekeys/width}}%
                {\pgfkeysvalueof{/tikz/bekeys/height}}%
                {1}%
                {0}%
            \fi
            \if\pgfkeysvalueof{/tikz/bekeys/direction}s
                {\pgfkeysvalueof{/tikz/bekeys/width}}%
                {\pgfkeysvalueof{/tikz/bekeys/height}}%
                {1}%
                {180}%
            \fi
            {\pgfkeysvalueof{/tikz/bekeys/color}}%
            {\pgfkeysvalueof{/tikz/bekeys/linestyle}}%
            {\pgfkeysvalueof{/tikz/bekeys/sbe}}%
            {\pgfkeysvalueof{/tikz/bekeys/angle}}%
            {\pgfkeysvalueof{/tikz/bekeys/fillgradient}}%
        }
    },
    /tikz/bekeys/drawer/.code n args={9}{%
        \tikzset{
            beshape,
            #6,
            #5,
            minimum height=
            \ifnum#3>0    % Flip width and height if #3 is not 0
            #1*\NODESIZE
            \else
            #2*\NODESIZE
            \fi
            ,minimum width=
            \ifnum#3>0
            #2*\NODESIZE
            \else
            #1*\NODESIZE
            \fi
            ,append after command={
                \pgfextra{
                    \let\bdr=\tikzlastnode%
                    \node[trapezium, line width = \NODETHICKNESS, minimum height=#1*\NODESIZE/3, minimum width=#2*\NODESIZE, trapezium stretches=true, rotate=#4, trapezium angle=70, inner sep=0.001mm, #5, #6, red] at (\bdr) (trap) {};

                    \node[rectangle, line width = \NODETHICKNESS, minimum height=#1*\NODESIZE/3, minimum width=#2*\NODESIZE, anchor=north, rotate=#4, #5, #6, green] at (trap.south) (r1) {};

                    \node[coordinate] at (trap.top left corner) (c1){};
                    \node[coordinate] at (trap.top right corner) (c2){};
                    \calcLength(c1,c2){distAB};

                    \node[rectangle, line width = \NODETHICKNESS, minimum height=#1*\NODESIZE/3, minimum width=\distAB, anchor=south, rotate=#4, #5, #6, red] at (trap.north) (r2) {};

					\draw[#5, #6, #9] (trap.top left corner) to (r2.north west) to (r2.north east) to (trap.top right corner) to (trap.bottom right corner) to (r1.south east) to (r1.south west) to (r1.north west) to cycle;

                }
            }
        }
    },
}
\pgfdeclareshape{couplershape}{	

	\inheritsavedanchors[from=basic] 

	\inheritanchor[from=basic]{center}
	\inheritanchor[from=basic]{mid}		
	\inheritanchor[from=basic]{base}	
	\inheritanchor[from=basic]{north}
	\inheritanchor[from=basic]{south}		
	\inheritanchor[from=basic]{west}		
	\inheritanchor[from=basic]{mid west}				
	\inheritanchor[from=basic]{base west}		
	\inheritanchor[from=basic]{north west}		
	\inheritanchor[from=basic]{south west}		
	\inheritanchor[from=basic]{east}
	\inheritanchor[from=basic]{mid east}
	\inheritanchor[from=basic]{base east}	
	\inheritanchor[from=basic]{north east}
	\inheritanchor[from=basic]{south east}
	\inheritanchor[from=basic]{west south west}%
	\inheritanchor[from=basic]{west north west}%
	\inheritanchor[from=basic]{east north east}%
	\inheritanchor[from=basic]{east south east}%
	\inheritanchor[from=basic]{north north west}%
	\inheritanchor[from=basic]{north north east}%
	\inheritanchor[from=basic]{south south west}%
	\inheritanchor[from=basic]{south south east}%		
	
	\inheritanchorborder[from=basic]	
	\inheritbackgroundpath[from=basic]
}

\pgfaddtoshape{couplershape}{
	\anchorlet{in}{center}		
	\anchorlet{in0}{center}
	\anchorlet{out}{center}
	\anchorlet{out0}{center}
	\anchorlet{c}{center}		
	\anchorlet{n}{north}
	\anchorlet{e}{east}
	\anchorlet{s}{south}
	\anchorlet{w}{west}			
	\anchorlet{se}{south east}
	\anchorlet{sw}{south west}
	\anchorlet{ne}{north east}
	\anchorlet{nw}{north west}
	\anchorlet{wsw}{west south west}
	\anchorlet{wnw}{west north west}
	\anchorlet{ene}{east north east}
	\anchorlet{ese}{east south east}
	\anchorlet{nnw}{north north west}
	\anchorlet{nne}{north north east}
	\anchorlet{ssw}{south south west}
	\anchorlet{sse}{south south east}
}

\tikzset{
	/tikz/couplerkeys/.cd,
	size/.initial=0.2,
	color/.initial=O,
	rotation/.initial=0,
	heightwidthratio/.initial=0.5,
	/tikz/coupler/.code={
		\pgfqkeys{/tikz/couplerkeys}{#1}%
		\tikzset{/tikz/couplerkeys/drawer/.expanded=%
			{\pgfkeysvalueof{/tikz/couplerkeys/size}}%
			{\pgfkeysvalueof{/tikz/couplerkeys/color}}%
			{\pgfkeysvalueof{/tikz/couplerkeys/rotation}}%
			{\pgfkeysvalueof{/tikz/couplerkeys/heightwidthratio}}%
		}
	},
	/tikz/couplerkeys/drawer/.code n args={4}{%
		\tikzset{
			couplershape,
			minimum height=#1*\NODESIZE
			\ifnum#3<1
				\ifnum#3>-1
					*#4
				\fi
			\fi
			,minimum width=#1*\NODESIZE
			\ifnum#3<91
				\ifnum#3>89
					*#4
				\fi
			\fi
			\ifnum#3<-89
				\ifnum#3>-91
					*#4
				\fi
			\fi
			,#2,
			append after command={
				\pgfextra{\let\bdr=\tikzlastnode%
				\node[ellipse, fill, #2, rotate=#3, outer sep = 0, minimum width=#1*\NODESIZE, minimum height=#1*#4*\NODESIZE] at (\bdr.center){};
				}
			}
		}
	},
}
\pgfdeclareshape{fibershape}{
	\savedmacro\direction{
		\edef\direction{\pgfkeysvalueof{/tikz/fiberkeys/direction}}%
	}
	\savedmacro\flip{
		\edef\flip{\pgfkeysvalueof{/tikz/fiberkeys/flip}}%
	}
	\saveddimen\minwidth{
		\pgfmathsetlength\pgf@x{\pgfshapeminwidth}%
	}
	\saveddimen\minheight{
		\pgfmathsetlength\pgf@x{\pgfshapeminheight}%
	}
	\inheritsavedanchors[from=basic] 
	
	\inheritanchor[from=basic]{center}
	\inheritanchor[from=basic]{mid}		
	\inheritanchor[from=basic]{base}	
	\inheritanchor[from=basic]{north}
	\inheritanchor[from=basic]{south}		
	\inheritanchor[from=basic]{west}		
	\inheritanchor[from=basic]{mid west}				
	\inheritanchor[from=basic]{base west}		
	\inheritanchor[from=basic]{north west}		
	\inheritanchor[from=basic]{south west}		
	\inheritanchor[from=basic]{east}
	\inheritanchor[from=basic]{mid east}
	\inheritanchor[from=basic]{base east}	
	\inheritanchor[from=basic]{north east}
	\inheritanchor[from=basic]{south east}
	\inheritanchor[from=basic]{west south west}%
	\inheritanchor[from=basic]{west north west}%
	\inheritanchor[from=basic]{east north east}%
	\inheritanchor[from=basic]{east south east}%
	\inheritanchor[from=basic]{north north west}%
	\inheritanchor[from=basic]{north north east}%
	\inheritanchor[from=basic]{south south west}%
	\inheritanchor[from=basic]{south south east}%		
	
	\inheritanchorborder[from=basic]	
	\inheritbackgroundpath[from=basic]

    \pgfutil@g@addto@macro\pgf@sh@s@fibershape{%
        \pgfutil@ifundefined{pgf@anchor@fibershape@in0}{%	If it is already defined, do not replace
	        \expandafter\xdef\csname pgf@anchor@fibershape@in0\endcsname{%
	            \noexpand\fibershape@port{0}% defined below
	        }%
	    }{}%
        \pgfutil@ifundefined{pgf@anchor@fibershape@in}{%	If it is already defined, do not replace
	        \expandafter\xdef\csname pgf@anchor@fibershape@in\endcsname{%
	            \noexpand\fibershape@port{0}% defined below
	        }%
	    }{}%
        \pgfutil@ifundefined{pgf@anchor@fibershape@out0}{%	If it is already defined, do not replace
	        \expandafter\xdef\csname pgf@anchor@fibershape@out0\endcsname{%
	            \noexpand\fibershape@port{1}% defined below
	        }%
	    }{}%
        \pgfutil@ifundefined{pgf@anchor@fibershape@out}{%	If it is already defined, do not replace
	        \expandafter\xdef\csname pgf@anchor@fibershape@out\endcsname{%
	            \noexpand\fibershape@port{1}% defined below
	        }%
	    }{}%
	}
}

\def\fibershape@port#1{%	#1 defines whether it is input or output
    \northeast	% Everything is defined wrt the northeast anchor

    \ifnum#1=0	% Port on the "left" side, the input
	    \if\direction\direce
			\pgf@x=-\pgf@x
	    	\if\flip\flipfalse
		    	\pgf@y=-\pgf@y
		    \else
		    	\pgf@y=\pgf@y
		    \fi
		\fi
	    \if\direction\direcw
			\pgf@x=\pgf@x
	    	\if\flip\flipfalse
		    	\pgf@y=-\pgf@y
		    \else
		    	\pgf@y=\pgf@y
		    \fi
		\fi
	    \if\direction\direcn
			\pgf@y=-\pgf@y
	    	\if\flip\flipfalse
		    	\pgf@x=-\pgf@x
		    \else
		    	\pgf@x=\pgf@x
		    \fi
		\fi
	    \if\direction\direcs
			\pgf@y=\pgf@y
	    	\if\flip\flipfalse
		    	\pgf@x=-\pgf@x
		    \else
		    	\pgf@x=\pgf@x
		    \fi
		\fi
	\else	% "Right" port, the output
	    \if\direction\direce
			\pgf@x=\pgf@x
	    	\if\flip\flipfalse
		    	\pgf@y=-\pgf@y
		    \else
		    	\pgf@y=\pgf@y
		    \fi
		\fi
	    \if\direction\direcw
			\pgf@x=-\pgf@x
	    	\if\flip\flipfalse
		    	\pgf@y=-\pgf@y
		    \else
		    	\pgf@y=\pgf@y
		    \fi
		\fi
	    \if\direction\direcn
			\pgf@y=\pgf@y
	    	\if\flip\flipfalse
		    	\pgf@x=-\pgf@x
		    \else
		    	\pgf@x=\pgf@x
		    \fi
		\fi
	    \if\direction\direcs
			\pgf@y=-\pgf@y
	    	\if\flip\flipfalse
		    	\pgf@x=-\pgf@x
		    \else
		    	\pgf@x=\pgf@x
		    \fi
		\fi
	\fi
}

\pgfaddtoshape{fibershape}{
	\anchorlet{c}{center}		
	\anchorlet{n}{north}
	\anchorlet{e}{east}
	\anchorlet{s}{south}
	\anchorlet{w}{west}			
	\anchorlet{se}{south east}
	\anchorlet{sw}{south west}
	\anchorlet{ne}{north east}
	\anchorlet{nw}{north west}
	\anchorlet{wsw}{west south west}
	\anchorlet{wnw}{west north west}
	\anchorlet{ene}{east north east}
	\anchorlet{ese}{east south east}
	\anchorlet{nnw}{north north west}
	\anchorlet{nne}{north north east}
	\anchorlet{ssw}{south south west}
	\anchorlet{sse}{south south east}
}

\tikzset{
	/tikz/fiberkeys/.cd,
	size/.initial=1,
	color/.initial=C0,
	direction/.initial=e,
	linestyle/.initial={linestyle},
	flip/.initial={0},
	drawbase/.initial={1},
	/tikz/fiber/.code={
		\pgfqkeys{/tikz/fiberkeys}{#1}%
		\tikzset{/tikz/fiberkeys/drawer/.expanded=%
			{\pgfkeysvalueof{/tikz/fiberkeys/direction}}%
			{\pgfkeysvalueof{/tikz/fiberkeys/size}}%
			{\pgfkeysvalueof{/tikz/fiberkeys/color}}%
			{\pgfkeysvalueof{/tikz/fiberkeys/linestyle}}%
			\if\pgfkeysvalueof{/tikz/fiberkeys/direction}e
				{a}%
				\if\pgfkeysvalueof{/tikz/fiberkeys/flip}0
					{south}%
				\else
					{north}%
				\fi
				{\pgfkeysvalueof{/tikz/fiberkeys/size}}% Width
				{\pgfkeysvalueof{/tikz/fiberkeys/size} * 0.5}% Height
			\fi
			\if\pgfkeysvalueof{/tikz/fiberkeys/direction}w
				{a}%
				\if\pgfkeysvalueof{/tikz/fiberkeys/flip}0
					{south}%
				\else
					{north}%
				\fi
				{\pgfkeysvalueof{/tikz/fiberkeys/size}}% Width
				{\pgfkeysvalueof{/tikz/fiberkeys/size} * 0.5}% Height
			\fi
			\if\pgfkeysvalueof{/tikz/fiberkeys/direction}n
				{b}%
				\if\pgfkeysvalueof{/tikz/fiberkeys/flip}0
					{west}%
				\else
					{east}%
				\fi
				{\pgfkeysvalueof{/tikz/fiberkeys/size} * 0.5}% Width
				{\pgfkeysvalueof{/tikz/fiberkeys/size}}% Height
			\fi
			\if\pgfkeysvalueof{/tikz/fiberkeys/direction}s
				{b}%
				\if\pgfkeysvalueof{/tikz/fiberkeys/flip}0
					{west}%
				\else
					{east}%
				\fi
				{\pgfkeysvalueof{/tikz/fiberkeys/size} * 0.5}% Width
				{\pgfkeysvalueof{/tikz/fiberkeys/size}}% Height
			\fi
			{\pgfkeysvalueof{/tikz/fiberkeys/drawbase}}%
		}
	},
	/tikz/fiberkeys/drawer/.code n args={9}{%
		\tikzset{
			fibershape,
			minimum width=#7*\NODESIZE,
			minimum height=#8*\NODESIZE,
			append after command={
				\pgfextra{\let\bdr=\tikzlastnode%
				\if#5a	% Variant a: East-West
					\ifnum#9>0
						\draw[#3, #4] (\bdr.#6 west) to (\bdr.#6 east) {};
					\fi
					\node[draw=#3, #4, circle, minimum size=#2*0.5*\NODESIZE, anchor=#6] at ([xshift=-0.1*#2*\NODESIZE]\bdr.#6) () {};
					\node[draw=#3, #4, circle, minimum size=#2*0.5*\NODESIZE, anchor=#6] at (\bdr.#6) () {};
					\node[draw=#3, #4, circle, minimum size=#2*0.5*\NODESIZE, anchor=#6] at ([xshift=0.1*#2*\NODESIZE]\bdr.#6) () {};
				\fi
				\if#5b	% Variant b: North-South
					\ifnum#9>0
						\draw[#3, #4] (\bdr.north #6) to (\bdr.south #6) {};
					\fi
					\node[draw=#3, #4, circle, minimum size=#2*0.5*\NODESIZE, anchor=#6] at ([yshift=0.1*#2*\NODESIZE]\bdr.#6) () {};
					\node[draw=#3, #4, circle, minimum size=#2*0.5*\NODESIZE, anchor=#6] at (\bdr.#6) () {};
					\node[draw=#3, #4, circle, minimum size=#2*0.5*\NODESIZE, anchor=#6] at ([yshift=-0.1*#2*\NODESIZE]\bdr.#6) () {};
				\fi
				}
			}
		}
	},
}
\newcount\portcount

\pgfdeclareshape{fiberswitchshape}{
    \savedmacro\nin{
        \edef\nin{\pgfkeysvalueof{/tikz/fiberswitchkeys/nin}}%
    }
    \savedmacro\nout{
        \edef\nout{\pgfkeysvalueof{/tikz/fiberswitchkeys/nout}}%
    }
    \savedmacro\direction{
        \edef\direction{\pgfkeysvalueof{/tikz/fiberswitchkeys/direction}}%
    }
    \inheritsavedanchors[from=basic]

    \inheritanchor[from=basic]{center}
    \inheritanchor[from=basic]{mid}
    \inheritanchor[from=basic]{base}
    \inheritanchor[from=basic]{north}
    \inheritanchor[from=basic]{south}
    \inheritanchor[from=basic]{west}
    \inheritanchor[from=basic]{mid west}
    \inheritanchor[from=basic]{base west}
    \inheritanchor[from=basic]{north west}
    \inheritanchor[from=basic]{south west}
    \inheritanchor[from=basic]{east}
    \inheritanchor[from=basic]{mid east}
    \inheritanchor[from=basic]{base east}
    \inheritanchor[from=basic]{north east}
    \inheritanchor[from=basic]{south east}
    \inheritanchor[from=basic]{west south west}%
    \inheritanchor[from=basic]{west north west}%
    \inheritanchor[from=basic]{east north east}%
    \inheritanchor[from=basic]{east south east}%
    \inheritanchor[from=basic]{north north west}%
    \inheritanchor[from=basic]{north north east}%
    \inheritanchor[from=basic]{south south west}%
    \inheritanchor[from=basic]{south south east}%		

    \inheritanchorborder[from=basic]
    \inheritbackgroundpath[from=basic]

    \pgfutil@g@addto@macro\pgf@sh@s@fiberswitchshape{%
        \pgfmathsetcount{\portcount}{0}
        \pgfmathloop%
        \ifnum\the\portcount<\nin%	Loop over the amount of ports
        \pgfutil@ifundefined{pgf@anchor@fiberswitchshape@in\the\portcount}{%	If it is already defined, do not replace
            \expandafter\xdef\csname pgf@anchor@fiberswitchshape@in\the\portcount\endcsname{%
                \noexpand\fiberswitchshape@port[\the\portcount]{0}% defined below
            }%
        }{}%
        \ifnum\the\portcount=0%	Make a general in anchor 
            \pgfutil@ifundefined{pgf@anchor@fiberswitchshape@in}{%
                \expandafter\xdef\csname pgf@anchor@fiberswitchshape@in\endcsname{%
                    \noexpand\fiberswitchshape@port[\the\portcount]{0}% defined below
                }%
            }{}%
        \fi
        \pgfmathaddtocount{\portcount}{1}	% Portcount += 1
        \repeatpgfmathloop%					% GOTO \pgfmathloop
        \pgfmathsetcount{\portcount}{0}
        \pgfmathloop%	        
        \ifnum\the\portcount<\nout% Add output ports, out0, out1, out2, ... up to \nout
        \pgfutil@ifundefined{pgf@anchor@fiberswitchshape@out\the\portcount}{%
            \expandafter\xdef\csname pgf@anchor@fiberswitchshape@out\the\portcount\endcsname{%
                \noexpand\fiberswitchshape@port[\the\portcount]{1}% defined below
            }%
        }{}%
        \ifnum\the\portcount=0%	Make a general out anchor 
            \pgfutil@ifundefined{pgf@anchor@fiberswitchshape@out}{%
                \expandafter\xdef\csname pgf@anchor@fiberswitchshape@out\endcsname{%
                    \noexpand\fiberswitchshape@port[\the\portcount]{1}% defined below
                }%
            }{}%
        \fi
        \pgfmathaddtocount{\portcount}{1}	% Portcount += 1
        \repeatpgfmathloop%					% GOTO \pgfmathloop
    }
}

\def\fiberswitchshape@port[#1]#2{%	#1 is the port number, #2 defines whether it is input or output
    \northeast \pgf@xa=\pgf@x \pgf@ya=\pgf@y% Store northeast in xa and ya
    \southwest \pgf@xb=\pgf@x \pgf@yb=\pgf@y% Store southwest in xb and yb

    \ifnum#2=0	% Input ports
        \if\direction\direce	% Direction E is the normal direction since most setup diagrams go from left to right
            \pgf@x=\pgf@xb
            \pgf@yc=\pgf@ya \advance\pgf@yc by -\pgf@yb	% yc = ya - yb
            \pgfmathsetlength{\pgf@y}{\pgf@ya-(#1 + 0.5)*(\pgf@yc/\nin)}%
        \fi
        \if\direction\direcw
            \pgf@x=\pgf@xa
            \pgf@yc=\pgf@ya \advance\pgf@yc by -\pgf@yb	% yc = ya - yb
            \pgfmathsetlength{\pgf@y}{\pgf@ya-(#1 + 0.5)*(\pgf@yc/\nin)}%
        \fi
        \if\direction\direcn
            \pgf@y=\pgf@yb
            \pgf@xc=\pgf@xa \advance\pgf@xc by -\pgf@xb	% xc = xa - xb
            \pgfmathsetlength{\pgf@x}{\pgf@xb+(#1 + 0.5)*(\pgf@xc/\nin)}%
        \fi
        \if\direction\direcs
            \pgf@y=\pgf@ya
            \pgf@xc=\pgf@xa \advance\pgf@xc by -\pgf@xb	% xc = xa - xb
            \pgfmathsetlength{\pgf@x}{\pgf@xb+(#1 + 0.5)*(\pgf@xc/\nin)}%
        \fi
    \else	% Output ports
        \if\direction\direce	% Direction E is the normal direction since most setup diagrams go from left to right
            \pgf@x=\pgf@xa
            \pgf@yc=\pgf@ya \advance\pgf@yc by -\pgf@yb	% yc = ya - yb
            \pgfmathsetlength{\pgf@y}{\pgf@ya-(#1 + 0.5)*(\pgf@yc/\nout)}%
        \fi
        \if\direction\direcw
            \pgf@x=\pgf@xb
            \pgf@yc=\pgf@ya \advance\pgf@yc by -\pgf@yb	% yc = ya - yb
            \pgfmathsetlength{\pgf@y}{\pgf@ya-(#1 + 0.5)*(\pgf@yc/\nout)}%
        \fi
        \if\direction\direcn
            \pgf@y=\pgf@ya
            \pgf@xc=\pgf@xa \advance\pgf@xc by -\pgf@xb	% xc = xa - xb
            \pgfmathsetlength{\pgf@x}{\pgf@xb+(#1 + 0.5)*(\pgf@xc/\nout)}%
        \fi
        \if\direction\direcs
            \pgf@y=\pgf@yb
            \pgf@xc=\pgf@xa \advance\pgf@xc by -\pgf@xb	% xc = xa - xb
            \pgfmathsetlength{\pgf@x}{\pgf@xb+(#1 + 0.5)*(\pgf@xc/\nout)}%
        \fi
    \fi
}

\pgfaddtoshape{fiberswitchshape}{
    \anchorlet{c}{center}
    \anchorlet{n}{north}
    \anchorlet{e}{east}
    \anchorlet{s}{south}
    \anchorlet{w}{west}
    \anchorlet{se}{south east}
    \anchorlet{sw}{south west}
    \anchorlet{ne}{north east}
    \anchorlet{nw}{north west}
    \anchorlet{wsw}{west south west}
    \anchorlet{wnw}{west north west}
    \anchorlet{ene}{east north east}
    \anchorlet{ese}{east south east}
    \anchorlet{nnw}{north north west}
    \anchorlet{nne}{north north east}
    \anchorlet{ssw}{south south west}
    \anchorlet{sse}{south south east}
}

\tikzset{
/tikz/fiberswitchkeys/.cd,
size/.initial=1,
color/.initial=O,
direction/.initial=e,
linestyle/.initial={linestyle, inner sep=0.5mm},
nin/.initial=1,	% Number of inputs is an option, but for now it is hardcoded for 1 input
nout/.initial=3,
fillgradient/.initial=,
drawarrow/.initial=1,
/tikz/fiberswitch/.code={
\pgfqkeys{/tikz/fiberswitchkeys}{#1}%
\tikzset{/tikz/fiberswitchkeys/drawer/.expanded=%
    {\pgfkeysvalueof{/tikz/fiberswitchkeys/size}}%
    {\pgfkeysvalueof{/tikz/fiberswitchkeys/color}}%
    {\pgfkeysvalueof{/tikz/fiberswitchkeys/linestyle}}%
    {\pgfkeysvalueof{/tikz/fiberswitchkeys/nout}}%
\if\pgfkeysvalueof{/tikz/fiberswitchkeys/direction}e
    {0}%
\fi
\if\pgfkeysvalueof{/tikz/fiberswitchkeys/direction}w
    {0}%
\fi
\if\pgfkeysvalueof{/tikz/fiberswitchkeys/direction}n
    {1}%
\fi
\if\pgfkeysvalueof{/tikz/fiberswitchkeys/direction}s
    {1}%
\fi
{\pgfkeysvalueof{/tikz/fiberswitchkeys/direction}}%
{\pgfkeysvalueof{/tikz/fiberswitchkeys/fillgradient}}%
{\pgfkeysvalueof{/tikz/fiberswitchkeys/drawarrow}}%
}
},
/tikz/fiberswitchkeys/drawer/.code n args={8}{%
        \tikzset{
            fiberswitchshape,
            draw,
            minimum height = #1*\NODESIZE,
            minimum width = #1*\NODESIZE,
            #2,
            #3,
            append after command={
                    \pgfextra{\let\bdr=\tikzlastnode%
                        \ifnum#5>0
                            \node[coordinate] at ($(\bdr.in)!0.25!(\bdr.out0 -| \bdr.in)$) (circlein){};
                            \foreach \n [evaluate=\n as \nport using int(\n-1)] in {1,...,#4}{
                                    \node[coordinate] at ($(\bdr.out\nport)!0.25!(\bdr.in -| \bdr.out\nport)$) (circleout\nport){};
                                }
                        \else
                            \node[coordinate] at ($(\bdr.in)!0.25!(\bdr.out0 |- \bdr.in)$) (circlein){};
                            \foreach \n [evaluate=\n as \nport using int(\n-1)] in {1,...,#4}{
                                    \node[coordinate] at ($(\bdr.out\nport)!0.25!(\bdr.in |- \bdr.out\nport)$) (circleout\nport){};
                                }
                        \fi

                        \node[rectangle, draw, fit=(\bdr.nw)(\bdr.se), #7] {}; % Needed for gradient fill

                        \draw[---, #2, #3, fill] (\bdr.in) to (circlein) circle (0.05);
                        \foreach \n [evaluate=\n as \nport using int(\n-1)] in {1,...,#4}{
                                \draw[---, #2, #3, fill] (\bdr.out\nport) to (circleout\nport) circle (0.05);
                            }

                        \tikzmath{
                            int \nportmax;
                            \nportmax = int(#4-1);
                        }
                        \draw[---, #2, #3] (circlein) to (circleout\nportmax){};
                        \ifnum#8>0
                        \if#6e
                            \draw[-->, #2, #3, looseness=0.8] ($(circlein)!0.6!(circleout0)$) to [out=-60, in=60]($(circlein)!0.6!(circleout\nportmax)$) {};
                        \fi
                        \if#6w
                            \draw[-->, #2, #3, looseness=0.8] ($(circlein)!0.6!(circleout0)$) to [out=-120, in=120]($(circlein)!0.6!(circleout\nportmax)$) {};
                        \fi
                        \if#6s
                            \draw[-->, #2, #3, looseness=0.8] ($(circlein)!0.6!(circleout0)$) to [out=-30, in=-150]($(circlein)!0.6!(circleout\nportmax)$) {};
                        \fi
                        \if#6n
                            \draw[-->, #2, #3, looseness=0.8] ($(circlein)!0.6!(circleout0)$) to [out=30, in=150]($(circlein)!0.6!(circleout\nportmax)$) {};
                        \fi
                        \fi

                    }
                }
        }
    },
}
\pgfdeclareshape{filtershape}{
	\savedmacro\direction{
		\edef\direction{\pgfkeysvalueof{/tikz/filterkeys/direction}}%
	}
	\saveddimen\minwidth{
		\pgfmathsetlength\pgf@x{\pgfshapeminwidth}%
	}
	\saveddimen\minheight{
		\pgfmathsetlength\pgf@x{\pgfshapeminheight}%
	}
	\inheritsavedanchors[from=basic]

	\inheritanchor[from=basic]{center}
	\inheritanchor[from=basic]{mid}
	\inheritanchor[from=basic]{base}
	\inheritanchor[from=basic]{north}
	\inheritanchor[from=basic]{south}
	\inheritanchor[from=basic]{west}
	\inheritanchor[from=basic]{mid west}
	\inheritanchor[from=basic]{base west}
	\inheritanchor[from=basic]{north west}
	\inheritanchor[from=basic]{south west}
	\inheritanchor[from=basic]{east}
	\inheritanchor[from=basic]{mid east}
	\inheritanchor[from=basic]{base east}
	\inheritanchor[from=basic]{north east}
	\inheritanchor[from=basic]{south east}
	\inheritanchor[from=basic]{west south west}%
	\inheritanchor[from=basic]{west north west}%
	\inheritanchor[from=basic]{east north east}%
	\inheritanchor[from=basic]{east south east}%
	\inheritanchor[from=basic]{north north west}%
	\inheritanchor[from=basic]{north north east}%
	\inheritanchor[from=basic]{south south west}%
	\inheritanchor[from=basic]{south south east}%		

	\inheritanchorborder[from=basic]
	\inheritbackgroundpath[from=basic]

	\pgfutil@g@addto@macro\pgf@sh@s@filtershape{%
		\pgfutil@ifundefined{pgf@anchor@filtershape@in0}{%	If it is already defined, do not replace
			\expandafter\xdef\csname pgf@anchor@filtershape@in0\endcsname{%
				\noexpand\filtershape@port{0}% defined below
			}%
		}{}%
		\pgfutil@ifundefined{pgf@anchor@filtershape@in}{%	If it is already defined, do not replace
			\expandafter\xdef\csname pgf@anchor@filtershape@in\endcsname{%
				\noexpand\filtershape@port{0}% defined below
			}%
		}{}%
		\pgfutil@ifundefined{pgf@anchor@filtershape@out0}{%	If it is already defined, do not replace
			\expandafter\xdef\csname pgf@anchor@filtershape@out0\endcsname{%
				\noexpand\filtershape@port{1}% defined below
			}%
		}{}%
		\pgfutil@ifundefined{pgf@anchor@filtershape@out}{%	If it is already defined, do not replace
			\expandafter\xdef\csname pgf@anchor@filtershape@out\endcsname{%
				\noexpand\filtershape@port{1}% defined below
			}%
		}{}%
	}
}

\def\filtershape@port#1{%	#1 defines whether it is input or output
	\northeast	% Everything is defined wrt the northeast anchor

	\ifnum#1=0	% Port on the "left" side, the input
		\if\direction\direce
			\pgf@x=-\pgf@x
			\pgf@ya= \pgf@y
			\pgfmathsetlength{\pgf@y}{\pgf@ya-0.5*\minheight}%
		\fi
		\if\direction\direcw
			\pgf@x=\pgf@x
			\pgf@ya= \pgf@y
			\pgfmathsetlength{\pgf@y}{\pgf@ya-0.5*\minheight}%
		\fi
		\if\direction\direcn
			\pgf@y=-\pgf@y
			\pgf@xa=\pgf@x
			\pgfmathsetlength{\pgf@x}{\pgf@xa-0.5*\minwidth}%
		\fi
		\if\direction\direcs
			\pgf@y=\pgf@y
			\pgf@xa= \pgf@x
			\pgfmathsetlength{\pgf@x}{\pgf@xa-0.5*\minwidth}%
		\fi
	\else	% "Right" port, the output
		\if\direction\direce
			\pgf@x=\pgf@x
			\pgf@ya= \pgf@y
			\pgfmathsetlength{\pgf@y}{\pgf@ya-0.5*\minheight}%
		\fi
		\if\direction\direcw
			\pgf@x=-\pgf@x
			\pgf@ya= \pgf@y
			\pgfmathsetlength{\pgf@y}{\pgf@ya-0.5*\minheight}%
		\fi
		\if\direction\direcn
			\pgf@y=\pgf@y
			\pgf@xa= \pgf@x
			\pgfmathsetlength{\pgf@x}{\pgf@xa-0.5*\minwidth}%
		\fi
		\if\direction\direcs
			\pgf@y=-\pgf@y
			\pgf@xa= \pgf@x
			\pgfmathsetlength{\pgf@x}{\pgf@xa-0.5*\minwidth}%
		\fi
	\fi
}

\pgfaddtoshape{filtershape}{
	\anchorlet{c}{center}
	\anchorlet{n}{north}
	\anchorlet{e}{east}
	\anchorlet{s}{south}
	\anchorlet{w}{west}
	\anchorlet{se}{south east}
	\anchorlet{sw}{south west}
	\anchorlet{ne}{north east}
	\anchorlet{nw}{north west}
	\anchorlet{wsw}{west south west}
	\anchorlet{wnw}{west north west}
	\anchorlet{ene}{east north east}
	\anchorlet{ese}{east south east}
	\anchorlet{nnw}{north north west}
	\anchorlet{nne}{north north east}
	\anchorlet{ssw}{south south west}
	\anchorlet{sse}{south south east}
}

\pgfmathsetmacro{\WSSSINEHEIGHT}{0.06}

\tikzset{
	/tikz/filterkeys/.cd,
	size/.initial=0.5,
	color/.initial=O,
	direction/.initial=e,
	linestyle/.initial={linestyle},
	fillgradient/.initial=O,
	/tikz/filter/.code={
			\pgfqkeys{/tikz/filterkeys}{#1}%
			\tikzset{/tikz/filterkeys/drawer/.expanded=%
					{\pgfkeysvalueof{/tikz/filterkeys/direction}}%
					{\pgfkeysvalueof{/tikz/filterkeys/size}}%
					{\pgfkeysvalueof{/tikz/filterkeys/color}}%
					{\pgfkeysvalueof{/tikz/filterkeys/linestyle}}%
					{\pgfkeysvalueof{/tikz/filterkeys/fillgradient}}%
			}
		},
	/tikz/filterkeys/drawer/.code n args={5}{%
			\tikzset{
				filtershape,
				minimum height=#2*\NODESIZE,
				minimum width=#2*\NODESIZE,
				#3,
				#4,
				draw,
				append after command={
						\pgfextra{\let\bdr=\tikzlastnode%
							\node[#5, fit=(\bdr.nw)(\bdr.se)] (boxgradient){};

							\node[coordinate] at (\bdr.wnw -| \bdr.nnw) (hl){};
							\node[coordinate] at (\bdr.ene -| \bdr.nne) (hr){};
							\draw[#3,---, rounded corners = 0] (hl) sin ($(hl)!0.25!(hr) + (0,0.002*#2*\NODESIZE)$) cos ($(hl)!0.5!(hr)$) sin ($(hl)!0.75!(hr) + (0,-0.002*#2*\NODESIZE)$) cos (hr);

							\node[coordinate] at (\bdr.w -| \bdr.nnw) (ml){};
							\node[coordinate] at (\bdr.e -| \bdr.nne) (mr){};
							\draw[#3,---, rounded corners = 0] (ml) sin ($(ml)!0.25!(mr) + (0,0.002*#2*\NODESIZE)$) cos ($(ml)!0.5!(mr)$) sin ($(ml)!0.75!(mr) + (0,-0.002*#2*\NODESIZE)$) cos (mr);

							\node[coordinate] at (\bdr.wsw -| \bdr.nnw) (ll){};
							\node[coordinate] at (\bdr.ese -| \bdr.nne) (lr){};
							\draw[#3,---, rounded corners = 0] (ll) sin ($(ll)!0.25!(lr) + (0,0.002*#2*\NODESIZE)$) cos ($(ll)!0.5!(lr)$) sin ($(ll)!0.75!(lr) + (0,-0.002*#2*\NODESIZE)$) cos (lr);

							\draw[#3, ---] ($(hl)!0.25!(hr) - (0,0.002*#2*\NODESIZE)$) -- ($(hl)!0.75!(hr) + (0,0.002*#2*\NODESIZE)$);
							\draw[#3, ---] ($(ll)!0.25!(lr) - (0,0.002*#2*\NODESIZE)$) -- ($(ll)!0.75!(lr) + (0,0.002*#2*\NODESIZE)$);
						}
					}
			}
		},
}
\newcount\portcount

\pgfdeclareshape{polswitchshape}{
    \savedmacro\nin{
        \edef\nin{\pgfkeysvalueof{/tikz/polswitchkeys/nin}}%
    }
    \savedmacro\nout{
        \edef\nout{\pgfkeysvalueof{/tikz/polswitchkeys/nout}}%
    }
    \savedmacro\direction{
        \edef\direction{\pgfkeysvalueof{/tikz/polswitchkeys/direction}}%
    }
    \inheritsavedanchors[from=basic]

    \inheritanchor[from=basic]{center}
    \inheritanchor[from=basic]{mid}
    \inheritanchor[from=basic]{base}
    \inheritanchor[from=basic]{north}
    \inheritanchor[from=basic]{south}
    \inheritanchor[from=basic]{west}
    \inheritanchor[from=basic]{mid west}
    \inheritanchor[from=basic]{base west}
    \inheritanchor[from=basic]{north west}
    \inheritanchor[from=basic]{south west}
    \inheritanchor[from=basic]{east}
    \inheritanchor[from=basic]{mid east}
    \inheritanchor[from=basic]{base east}
    \inheritanchor[from=basic]{north east}
    \inheritanchor[from=basic]{south east}
    \inheritanchor[from=basic]{west south west}%
    \inheritanchor[from=basic]{west north west}%
    \inheritanchor[from=basic]{east north east}%
    \inheritanchor[from=basic]{east south east}%
    \inheritanchor[from=basic]{north north west}%
    \inheritanchor[from=basic]{north north east}%
    \inheritanchor[from=basic]{south south west}%
    \inheritanchor[from=basic]{south south east}%		

    \inheritanchorborder[from=basic]
    \inheritbackgroundpath[from=basic]

    \pgfutil@g@addto@macro\pgf@sh@s@polswitchshape{%
        \pgfmathsetcount{\portcount}{0}
        \pgfmathloop%
        \ifnum\the\portcount<\nin%	Loop over the amount of ports
        \pgfutil@ifundefined{pgf@anchor@polswitchshape@in\the\portcount}{%	If it is already defined, do not replace
            \expandafter\xdef\csname pgf@anchor@polswitchshape@in\the\portcount\endcsname{%
                \noexpand\polswitchshape@port[\the\portcount]{0}% defined below
            }%
        }{}%
        \ifnum\the\portcount=0%	Make a general in anchor 
            \pgfutil@ifundefined{pgf@anchor@polswitchshape@in}{%
                \expandafter\xdef\csname pgf@anchor@polswitchshape@in\endcsname{%
                    \noexpand\polswitchshape@port[\the\portcount]{0}% defined below
                }%
            }{}%
        \fi
        \pgfmathaddtocount{\portcount}{1}	% Portcount += 1
        \repeatpgfmathloop%					% GOTO \pgfmathloop
        \pgfmathsetcount{\portcount}{0}
        \pgfmathloop%	        
        \ifnum\the\portcount<\nout% Add output ports, out0, out1, out2, ... up to \nout
        \pgfutil@ifundefined{pgf@anchor@polswitchshape@out\the\portcount}{%
            \expandafter\xdef\csname pgf@anchor@polswitchshape@out\the\portcount\endcsname{%
                \noexpand\polswitchshape@port[\the\portcount]{1}% defined below
            }%
        }{}%
        \ifnum\the\portcount=0%	Make a general out anchor 
            \pgfutil@ifundefined{pgf@anchor@polswitchshape@out}{%
                \expandafter\xdef\csname pgf@anchor@polswitchshape@out\endcsname{%
                    \noexpand\polswitchshape@port[\the\portcount]{1}% defined below
                }%
            }{}%
        \fi
        \pgfmathaddtocount{\portcount}{1}	% Portcount += 1
        \repeatpgfmathloop%					% GOTO \pgfmathloop
    }
}

\def\polswitchshape@port[#1]#2{%	#1 is the port number, #2 defines whether it is input or output
    \northeast \pgf@xa=\pgf@x \pgf@ya=\pgf@y% Store northeast in xa and ya
    \southwest \pgf@xb=\pgf@x \pgf@yb=\pgf@y% Store southwest in xb and yb

    \ifnum#2=0	% Input ports
        \if\direction\direce	% Direction E is the normal direction since most setup diagrams go from left to right
            \pgf@x=\pgf@xb
            \pgf@yc=\pgf@ya \advance\pgf@yc by -\pgf@yb	% yc = ya - yb
            \pgfmathsetlength{\pgf@y}{\pgf@ya-(#1 + 0.5)*(\pgf@yc/\nin)}%
        \fi
        \if\direction\direcw
            \pgf@x=\pgf@xa
            \pgf@yc=\pgf@ya \advance\pgf@yc by -\pgf@yb	% yc = ya - yb
            \pgfmathsetlength{\pgf@y}{\pgf@ya-(#1 + 0.5)*(\pgf@yc/\nin)}%
        \fi
        \if\direction\direcn
            \pgf@y=\pgf@yb
            \pgf@xc=\pgf@xa \advance\pgf@xc by -\pgf@xb	% xc = xa - xb
            \pgfmathsetlength{\pgf@x}{\pgf@xb+(#1 + 0.5)*(\pgf@xc/\nin)}%
        \fi
        \if\direction\direcs
            \pgf@y=\pgf@ya
            \pgf@xc=\pgf@xa \advance\pgf@xc by -\pgf@xb	% xc = xa - xb
            \pgfmathsetlength{\pgf@x}{\pgf@xb+(#1 + 0.5)*(\pgf@xc/\nin)}%
        \fi
    \else	% Output ports
        \if\direction\direce	% Direction E is the normal direction since most setup diagrams go from left to right
            \pgf@x=\pgf@xa
            \pgf@yc=\pgf@ya \advance\pgf@yc by -\pgf@yb	% yc = ya - yb
            \pgfmathsetlength{\pgf@y}{\pgf@ya-(#1 + 0.5)*(\pgf@yc/\nout)}%
        \fi
        \if\direction\direcw
            \pgf@x=\pgf@xb
            \pgf@yc=\pgf@ya \advance\pgf@yc by -\pgf@yb	% yc = ya - yb
            \pgfmathsetlength{\pgf@y}{\pgf@ya-(#1 + 0.5)*(\pgf@yc/\nout)}%
        \fi
        \if\direction\direcn
            \pgf@y=\pgf@ya
            \pgf@xc=\pgf@xa \advance\pgf@xc by -\pgf@xb	% xc = xa - xb
            \pgfmathsetlength{\pgf@x}{\pgf@xb+(#1 + 0.5)*(\pgf@xc/\nout)}%
        \fi
        \if\direction\direcs
            \pgf@y=\pgf@yb
            \pgf@xc=\pgf@xa \advance\pgf@xc by -\pgf@xb	% xc = xa - xb
            \pgfmathsetlength{\pgf@x}{\pgf@xb+(#1 + 0.5)*(\pgf@xc/\nout)}%
        \fi
    \fi
}

\pgfaddtoshape{polswitchshape}{
    \anchorlet{c}{center}
    \anchorlet{n}{north}
    \anchorlet{e}{east}
    \anchorlet{s}{south}
    \anchorlet{w}{west}
    \anchorlet{se}{south east}
    \anchorlet{sw}{south west}
    \anchorlet{ne}{north east}
    \anchorlet{nw}{north west}
    \anchorlet{wsw}{west south west}
    \anchorlet{wnw}{west north west}
    \anchorlet{ene}{east north east}
    \anchorlet{ese}{east south east}
    \anchorlet{nnw}{north north west}
    \anchorlet{nne}{north north east}
    \anchorlet{ssw}{south south west}
    \anchorlet{sse}{south south east}
}

\tikzset{
/tikz/polswitchkeys/.cd,
size/.initial=1,
color/.initial=O,
direction/.initial=e,
linestyle/.initial={linestyle, inner sep=0.5mm},
nin/.initial=1,	% Number of inputs is an option, but for now it is hardcoded for 1 input
nout/.initial=1, % Number of outputs is an option, but for now it is hardcoded for 1 output
fillgradient/.initial=,
/tikz/polswitch/.code={
\pgfqkeys{/tikz/polswitchkeys}{#1}%
\tikzset{/tikz/polswitchkeys/drawer/.expanded=%
    {\pgfkeysvalueof{/tikz/polswitchkeys/size}}%
    {\pgfkeysvalueof{/tikz/polswitchkeys/color}}%
    {\pgfkeysvalueof{/tikz/polswitchkeys/linestyle}}%
    {\pgfkeysvalueof{/tikz/polswitchkeys/nout}}%
\if\pgfkeysvalueof{/tikz/polswitchkeys/direction}e
    {0}%
\fi
\if\pgfkeysvalueof{/tikz/polswitchkeys/direction}w
    {0}%
\fi
\if\pgfkeysvalueof{/tikz/polswitchkeys/direction}n
    {1}%
\fi
\if\pgfkeysvalueof{/tikz/polswitchkeys/direction}s
    {1}%
\fi
{\pgfkeysvalueof{/tikz/polswitchkeys/direction}}%
{\pgfkeysvalueof{/tikz/polswitchkeys/fillgradient}}%
}
},
/tikz/polswitchkeys/drawer/.code n args={7}{%
        \tikzset{
            polswitchshape,
            draw,
            minimum height = #1*\NODESIZE,
            minimum width = #1*\NODESIZE,
            #2,
            #3,
            append after command={
                    \pgfextra{\let\bdr=\tikzlastnode%
                        \node[rectangle, draw, fit=(\bdr.nw)(\bdr.se), #7] {}; % Needed for gradient fill

                        \node[coordinate] at ($(\bdr.in)!0.25!(\bdr.out)$) (circlein){};
                        \node[coordinate] at ($(\bdr.in)!0.75!(\bdr.out)$) (circleout){};

                        \draw[---, #2, #3, fill] (\bdr.in) to (circlein) circle (0.05);
                        \draw[---, #2, #3, fill] (\bdr.out) to (circleout) circle (0.05);

                        \node[coordinate] at ($(\bdr.in)!0.6!(\bdr.out)$) (circlemiddle){};
                        \ifnum#5>0
                            \node[coordinate] at ($(circlemiddle)!0.5!(circlemiddle -| \bdr.e)$) (circletopcor){};
                            \node[coordinate] at ($(circlemiddle)!0.5!(circlemiddle -| \bdr.w)$) (circlebotcor){};
                        \else
                            \node[coordinate] at ($(circlemiddle)!0.5!(circlemiddle |- \bdr.n)$) (circletopcor){};
                            \node[coordinate] at ($(circlemiddle)!0.5!(circlemiddle |- \bdr.s)$) (circlebotcor){};
                        \fi

                        \node[draw, circle, #2, #3, minimum size=0.3*\FNODESIZE] at (circletopcor) (circletop){};
                        \node[draw, circle, #2, #3, minimum size=0.3*\FNODESIZE] at (circlebotcor) (circlebot){};
                        \draw[-->, #2, #3] (circletop.south) to (circletop.north){};
                        \draw[-->, #2, #3] (circlebot.west) to (circlebot.east){};

                        \if#6e
                            \draw[-->, #2, #3] ([xshift=-0.05*\FNODESIZE]circletop.south west) to [out=-120, in=120] ([xshift=-0.05*\FNODESIZE]circlebot.north west){};
                        \fi
                        \if#6w
                            \draw[-->, #2, #3] ([xshift=0.05*\FNODESIZE]circletop.south east) to [out=-60, in=60] ([xshift=0.05*\FNODESIZE]circlebot.north east){};
                        \fi
                        \if#6s
                            \draw[-->, #2, #3] ([yshift=0.05*\FNODESIZE]circletop.north west) to [out=150, in=30] ([yshift=0.05*\FNODESIZE]circlebot.north east){};
                        \fi
                        \if#6n
                            \draw[-->, #2, #3] ([yshift=-0.05*\FNODESIZE]circletop.south west) to [out=-150, in=-30] ([yshift=-0.05*\FNODESIZE]circlebot.south east){};
                        \fi

                    }
                }
        }
    },
}
\pgfdeclareshape{pdshape}{
	\savedmacro\direction{
		\edef\direction{\pgfkeysvalueof{/tikz/pdkeys/direction}}%
	}
	\saveddimen\minwidth{
		\pgfmathsetlength\pgf@x{\pgfshapeminwidth}%
	}
	\saveddimen\minheight{
		\pgfmathsetlength\pgf@x{\pgfshapeminheight}%
	}
	\inheritsavedanchors[from=basic]

	\inheritanchor[from=basic]{center}
	\inheritanchor[from=basic]{mid}
	\inheritanchor[from=basic]{base}
	\inheritanchor[from=basic]{north}
	\inheritanchor[from=basic]{south}
	\inheritanchor[from=basic]{west}
	\inheritanchor[from=basic]{mid west}
	\inheritanchor[from=basic]{base west}
	\inheritanchor[from=basic]{north west}
	\inheritanchor[from=basic]{south west}
	\inheritanchor[from=basic]{east}
	\inheritanchor[from=basic]{mid east}
	\inheritanchor[from=basic]{base east}
	\inheritanchor[from=basic]{north east}
	\inheritanchor[from=basic]{south east}
	\inheritanchor[from=basic]{west south west}%
	\inheritanchor[from=basic]{west north west}%
	\inheritanchor[from=basic]{east north east}%
	\inheritanchor[from=basic]{east south east}%
	\inheritanchor[from=basic]{north north west}%
	\inheritanchor[from=basic]{north north east}%
	\inheritanchor[from=basic]{south south west}%
	\inheritanchor[from=basic]{south south east}%		

	\inheritanchorborder[from=basic]
	\inheritbackgroundpath[from=basic]

	\pgfutil@g@addto@macro\pgf@sh@s@pdshape{%
		\pgfutil@ifundefined{pgf@anchor@pdshape@in0}{%	If it is already defined, do not replace
			\expandafter\xdef\csname pgf@anchor@pdshape@in0\endcsname{%
				\noexpand\pdshape@port{0}% defined below
			}%
		}{}%
		\pgfutil@ifundefined{pgf@anchor@pdshape@in}{%	If it is already defined, do not replace
			\expandafter\xdef\csname pgf@anchor@pdshape@in\endcsname{%
				\noexpand\pdshape@port{0}% defined below
			}%
		}{}%
		\pgfutil@ifundefined{pgf@anchor@pdshape@out0}{%	If it is already defined, do not replace
			\expandafter\xdef\csname pgf@anchor@pdshape@out0\endcsname{%
				\noexpand\pdshape@port{1}% defined below
			}%
		}{}%
		\pgfutil@ifundefined{pgf@anchor@pdshape@out}{%	If it is already defined, do not replace
			\expandafter\xdef\csname pgf@anchor@pdshape@out\endcsname{%
				\noexpand\pdshape@port{1}% defined below
			}%
		}{}%
	}
}

\def\pdshape@port#1{%	#1 defines whether it is input or output
	\northeast	% Everything is defined wrt the northeast anchor

	\ifnum#1=0	% Port on the "left" side, the input
		\if\direction\direce
			\pgf@x=-\pgf@x
			\pgf@ya= \pgf@y
			\pgfmathsetlength{\pgf@y}{\pgf@ya-0.5*\minheight}%
		\fi
		\if\direction\direcw
			\pgf@x=\pgf@x
			\pgf@ya= \pgf@y
			\pgfmathsetlength{\pgf@y}{\pgf@ya-0.5*\minheight}%
		\fi
		\if\direction\direcn
			\pgf@y=-\pgf@y
			\pgf@xa=\pgf@x
			\pgfmathsetlength{\pgf@x}{\pgf@xa-0.5*\minwidth}%
		\fi
		\if\direction\direcs
			\pgf@y=\pgf@y
			\pgf@xa= \pgf@x
			\pgfmathsetlength{\pgf@x}{\pgf@xa-0.5*\minwidth}%
		\fi
	\else	% "Right" port, the output
		\if\direction\direce
			\pgf@x=\pgf@x
			\pgf@ya= \pgf@y
			\pgfmathsetlength{\pgf@y}{\pgf@ya-0.5*\minheight}%
		\fi
		\if\direction\direcw
			\pgf@x=-\pgf@x
			\pgf@ya= \pgf@y
			\pgfmathsetlength{\pgf@y}{\pgf@ya-0.5*\minheight}%
		\fi
		\if\direction\direcn
			\pgf@y=\pgf@y
			\pgf@xa= \pgf@x
			\pgfmathsetlength{\pgf@x}{\pgf@xa-0.5*\minwidth}%
		\fi
		\if\direction\direcs
			\pgf@y=-\pgf@y
			\pgf@xa= \pgf@x
			\pgfmathsetlength{\pgf@x}{\pgf@xa-0.5*\minwidth}%
		\fi
	\fi
}

\pgfaddtoshape{pdshape}{
	\anchorlet{c}{center}
	\anchorlet{n}{north}
	\anchorlet{e}{east}
	\anchorlet{s}{south}
	\anchorlet{w}{west}
	\anchorlet{se}{south east}
	\anchorlet{sw}{south west}
	\anchorlet{ne}{north east}
	\anchorlet{nw}{north west}
	\anchorlet{wsw}{west south west}
	\anchorlet{wnw}{west north west}
	\anchorlet{ene}{east north east}
	\anchorlet{ese}{east south east}
	\anchorlet{nnw}{north north west}
	\anchorlet{nne}{north north east}
	\anchorlet{ssw}{south south west}
	\anchorlet{sse}{south south east}
}

\tikzset{
/tikz/pdkeys/.cd,
size/.initial=0.5,
color/.initial=EO,
direction/.initial=e,
linestyle/.initial={linestyle},
fillgradient/.initial=O,
/tikz/pd/.code={
		\pgfqkeys{/tikz/pdkeys}{#1}%
		\tikzset{/tikz/pdkeys/drawer/.expanded=%
				{\pgfkeysvalueof{/tikz/pdkeys/direction}}%
				{\pgfkeysvalueof{/tikz/pdkeys/size}}%
				{\pgfkeysvalueof{/tikz/pdkeys/color}}%
				{\pgfkeysvalueof{/tikz/pdkeys/linestyle}}%
				{\pgfkeysvalueof{/tikz/pdkeys/fillgradient}}%
		}
	},
/tikz/pdkeys/drawer/.code n args={5}{%
\tikzset{
pdshape,
minimum height=#2*\NODESIZE,
minimum width=#2*\NODESIZE,
#3,
#4,
draw,
append after command={
\pgfextra{\let\bdr=\tikzlastnode%
\node[#5, fit=(\bdr.nw)(\bdr.se)] (boxgradient){};
\draw[---,#3] ($(\bdr.s)!.1!(\bdr.n)$) to ($(\bdr.s)!.9!(\bdr.n)$);
\fill[#3] ({$(\bdr.s)!.3!(\bdr.n)$} -| {$(\bdr.w)!.3!(\bdr.e)$}) to ($(\bdr.s)!.7!(\bdr.n)$) to ({$(\bdr.s)!.3!(\bdr.n)$} -| {$(\bdr.w)!.7!(\bdr.e)$}) to cycle;
\draw[---,#3] ({$(\bdr.s)!.7!(\bdr.n)$} -| {$(\bdr.w)!.35!(\bdr.e)$}) to ({$(\bdr.s)!.7!(\bdr.n)$} -| {$(\bdr.w)!.65!(\bdr.e)$});
}
}
}
},
}
\pgfdeclareshape{pbsshape}{
	\savedmacro\direction{
		\edef\direction{\pgfkeysvalueof{/tikz/pbskeys/direction}}%
	}
	\saveddimen\minwidth{
		\pgfmathsetlength\pgf@x{\pgfshapeminwidth}%
	}
	\saveddimen\minheight{
		\pgfmathsetlength\pgf@x{\pgfshapeminheight}%
	}
	\savedmacro\nport{
		\edef\nport{\pgfkeysvalueof{/tikz/pbskeys/nport}}%
	}
	\inheritsavedanchors[from=basic]

	\inheritanchor[from=basic]{center}
	\inheritanchor[from=basic]{mid}
	\inheritanchor[from=basic]{base}
	\inheritanchor[from=basic]{north}
	\inheritanchor[from=basic]{south}
	\inheritanchor[from=basic]{west}
	\inheritanchor[from=basic]{mid west}
	\inheritanchor[from=basic]{base west}
	\inheritanchor[from=basic]{north west}
	\inheritanchor[from=basic]{south west}
	\inheritanchor[from=basic]{east}
	\inheritanchor[from=basic]{mid east}
	\inheritanchor[from=basic]{base east}
	\inheritanchor[from=basic]{north east}
	\inheritanchor[from=basic]{south east}
	\inheritanchor[from=basic]{west south west}%
	\inheritanchor[from=basic]{west north west}%
	\inheritanchor[from=basic]{east north east}%
	\inheritanchor[from=basic]{east south east}%
	\inheritanchor[from=basic]{north north west}%
	\inheritanchor[from=basic]{north north east}%
	\inheritanchor[from=basic]{south south west}%
	\inheritanchor[from=basic]{south south east}%

	\inheritanchorborder[from=basic]
	\inheritbackgroundpath[from=basic]

	\pgfutil@g@addto@macro\pgf@sh@s@pbsshape{%
		\pgfutil@ifundefined{pgf@anchor@pbsshape@in0}{%	If it is already defined, do not replace
			\expandafter\xdef\csname pgf@anchor@pbsshape@in0\endcsname{%
				\noexpand\pbsshape@port[0]{0}% defined below
			}%
		}{}%
		\pgfutil@ifundefined{pgf@anchor@pbsshape@in1}{%	If it is already defined, do not replace
			\expandafter\xdef\csname pgf@anchor@pbsshape@in1\endcsname{%
				\noexpand\pbsshape@port[1]{0}% defined below
			}%
		}{}%
		\pgfutil@ifundefined{pgf@anchor@pbsshape@in}{%	If it is already defined, do not replace
			\expandafter\xdef\csname pgf@anchor@pbsshape@in\endcsname{%
				\noexpand\pbsshape@port[0]{0}% defined below
			}%
		}{}%
		\pgfutil@ifundefined{pgf@anchor@pbsshape@out0}{%	If it is already defined, do not replace
			\expandafter\xdef\csname pgf@anchor@pbsshape@out0\endcsname{%
				\noexpand\pbsshape@port[0]{1}% defined below
			}%
		}{}%
		\pgfutil@ifundefined{pgf@anchor@pbsshape@out1}{%	If it is already defined, do not replace
			\expandafter\xdef\csname pgf@anchor@pbsshape@out1\endcsname{%
				\noexpand\pbsshape@port[1]{1}% defined below
			}%
		}{}%
		\pgfutil@ifundefined{pgf@anchor@pbsshape@out}{%	If it is already defined, do not replace
			\expandafter\xdef\csname pgf@anchor@pbsshape@out\endcsname{%
				\noexpand\pbsshape@port[0]{1}% defined below
			}%
		}{}%
		\pgfmathsetcount{\portcount}{0}
		\pgfmathloop%
		\ifnum\the\portcount<\nport%	Loop over the amount of ports
		\pgfutil@ifundefined{pgf@anchor@pbsshape@p\the\portcount}{%	If it is already defined, do not replace
			\expandafter\xdef\csname pgf@anchor@pbsshape@p\the\portcount\endcsname{%
				\noexpand\pbsshape@port[\the\portcount]{2}% defined below
			}%
		}{}%
		\pgfmathaddtocount{\portcount}{1}	% Portcount += 1
		\repeatpgfmathloop%					% GOTO \pgfmathloop
	}
}

\def\pbsshape@port[#1]#2{%	#1 is the port number, #2 defines whether it is input or output
	\northeast	% Everything is defined wrt the northeast anchor

	\ifnum#2=0	% Port on the "left" side, the input
		\if\direction\direce
			\ifnum#1=0
				\pgf@x=-\pgf@x
				\pgf@ya= \pgf@y
				\pgfmathsetlength{\pgf@y}{\pgf@ya-0.5*\minheight}%
			\else
				\pgf@x=0\pgf@x
				\pgf@ya= \pgf@y
				\pgfmathsetlength{\pgf@y}{\pgf@ya-\minheight}%
			\fi
		\fi
		\if\direction\direcw
			\ifnum#1=0
				\pgf@x=\pgf@x
				\pgf@ya= \pgf@y
				\pgfmathsetlength{\pgf@y}{\pgf@ya-0.5*\minheight}%
			\else
				\pgf@x=0\pgf@x
				\pgf@ya= \pgf@y
				\pgfmathsetlength{\pgf@y}{\pgf@ya}%
			\fi
		\fi
		\if\direction\direcn
			\ifnum#1=0
				\pgf@y=-\pgf@y
				\pgf@xa=\pgf@x
				\pgfmathsetlength{\pgf@x}{\pgf@xa-0.5*\minwidth}%
			\else
				\pgf@y=0\pgf@y
				\pgf@xa=\pgf@x
				\pgfmathsetlength{\pgf@x}{\pgf@xa-0*\minwidth}%
			\fi
		\fi
		\if\direction\direcs
			\ifnum#1=0
				\pgf@y=\pgf@y
				\pgf@xa= \pgf@x
				\pgfmathsetlength{\pgf@x}{\pgf@xa-0.5*\minwidth}%
			\else
				\pgf@y=0\pgf@y
				\pgf@xa= \pgf@x
				\pgfmathsetlength{\pgf@x}{\pgf@xa-1*\minwidth}%
			\fi
		\fi
	\else	% "Right" port, the output
		\if\direction\direce
			\ifnum#1=0
				\pgf@x=\pgf@x
				\pgf@ya= \pgf@y
				\pgfmathsetlength{\pgf@y}{\pgf@ya-0.5*\minheight}%
			\else
				\pgf@x=0\pgf@x
				\pgf@ya= \pgf@y
				\pgfmathsetlength{\pgf@y}{\pgf@ya}%
			\fi
		\fi
		\if\direction\direcw
			\ifnum#1=0
				\pgf@x=-\pgf@x
				\pgf@ya= \pgf@y
				\pgfmathsetlength{\pgf@y}{\pgf@ya-0.5*\minheight}%
			\else
				\pgf@x=0\pgf@x
				\pgf@ya= \pgf@y
				\pgfmathsetlength{\pgf@y}{\pgf@ya-\minheight}%
			\fi
		\fi
		\if\direction\direcn
			\ifnum#1=0
				\pgf@y=\pgf@y
				\pgf@xa= \pgf@x
				\pgfmathsetlength{\pgf@x}{\pgf@xa-0.5*\minwidth}%
			\else
				\pgf@y=0\pgf@y
				\pgf@xa= \pgf@x
				\pgfmathsetlength{\pgf@x}{\pgf@xa-1*\minwidth}%
			\fi
		\fi
		\if\direction\direcs
			\ifnum#1=0
				\pgf@y=-\pgf@y
				\pgf@xa= \pgf@x
				\pgfmathsetlength{\pgf@x}{\pgf@xa-0.5*\minwidth}%
			\else
				\pgf@y=0\pgf@y
				\pgf@xa= \pgf@x
				\pgfmathsetlength{\pgf@x}{\pgf@xa-0*\minwidth}%
			\fi
		\fi
	\fi

	\ifnum#2=2	%
		\northeast \pgf@xa=\pgf@x \pgf@ya=\pgf@y% Store northeast in xa and ya
		\southwest \pgf@xb=\pgf@x \pgf@yb=\pgf@y% Store southwest in xb and yb

		\pgf@xc=\pgf@xa \advance\pgf@xc by -\pgf@xb	% xc = xa - xb
		\pgfmathsetlength{\pgf@x}{\pgf@xa-(#1 + 0.5)*(\pgf@xc/\nport)}%
		\pgf@yc=\pgf@ya \advance\pgf@yc by -\pgf@yb	% yc = ya - yb
		\pgfmathsetlength{\pgf@y}{\pgf@ya-(#1 + 0.5)*(\pgf@yc/\nport)}%

		\if\direction\direce
			\pgf@x=-\pgf@x
		\fi
		\if\direction\direcw
			\pgf@x=-\pgf@x
		\fi
	\fi
}

\pgfaddtoshape{pbsshape}{
	\anchorlet{c}{center}
	\anchorlet{n}{north}
	\anchorlet{e}{east}
	\anchorlet{s}{south}
	\anchorlet{w}{west}
	\anchorlet{se}{south east}
	\anchorlet{sw}{south west}
	\anchorlet{ne}{north east}
	\anchorlet{nw}{north west}
	\anchorlet{wsw}{west south west}
	\anchorlet{wnw}{west north west}
	\anchorlet{ene}{east north east}
	\anchorlet{ese}{east south east}
	\anchorlet{nnw}{north north west}
	\anchorlet{nne}{north north east}
	\anchorlet{ssw}{south south west}
	\anchorlet{sse}{south south east}
}

\pgfmathsetmacro{\WSSSINEHEIGHT}{0.06}

\tikzset{
	/tikz/pbskeys/.cd,
	size/.initial=0.5,
	color/.initial=O,
	direction/.initial=e,
	linestyle/.initial={linestyle},
	nport/.initial=1,
	fillgradient/.initial=O,
	/tikz/pbs/.code={
			\pgfqkeys{/tikz/pbskeys}{#1}%
			\tikzset{/tikz/pbskeys/drawer/.expanded=%
					{\pgfkeysvalueof{/tikz/pbskeys/direction}}%
					{\pgfkeysvalueof{/tikz/pbskeys/size}}%
					{\pgfkeysvalueof{/tikz/pbskeys/color}}%
					{\pgfkeysvalueof{/tikz/pbskeys/linestyle}}%
					{\pgfkeysvalueof{/tikz/pbskeys/fillgradient}}%
			}
		},
	/tikz/pbskeys/drawer/.code n args={5}{%
			\tikzset{
				pbsshape,
				minimum height=#2*\NODESIZE,
				minimum width=#2*\NODESIZE,
				#3,
				#4,
				draw,
				append after command={
						\pgfextra{\let\bdr=\tikzlastnode%
							\node[#5, fit=(\bdr.nw)(\bdr.se)] (boxgradient){};

							\if#1e
								\draw[#3, ---] ($(\bdr.nw)!.01!(\bdr.se)$) to ($(\bdr.se)!.01!(\bdr.nw)$);
							\fi
							\if#1w
								\draw[#3, ---] ($(\bdr.nw)!.01!(\bdr.se)$) to ($(\bdr.se)!.01!(\bdr.nw)$);
							\fi
							\if#1n
								\draw[#3, ---] ($(\bdr.ne)!.01!(\bdr.sw)$) to ($(\bdr.sw)!.01!(\bdr.ne)$);
							\fi
							\if#1s
								\draw[#3, ---] ($(\bdr.ne)!.01!(\bdr.sw)$) to ($(\bdr.sw)!.01!(\bdr.ne)$);
							\fi

						}
					}
			}
		},
}

\ifdefined\fontchoice
\else
	\def\fontchoice{times} % By default choose times
\fi

\usepackage{pdftexcmds}

\ifnum\pdf@strcmp{\fontchoice}{firasans}=0 %
	\usepackage[book]{FiraSans} %% option 'sfdefault' activates Fira Sans as the default text font
	\usepackage[T1]{fontenc}
	
	\tikzset{every picture/.style={/utils/exec={\sffamily}}}
\fi

\ifnum\pdf@strcmp{\fontchoice}{times}=0 %
	\usepackage{times}
\fi

\ifnum\pdf@strcmp{\fontchoice}{timesnewroman}=0 %
	\usepackage{mathptmx}
	\usepackage[T1]{fontenc}
\fi

\ifnum\pdf@strcmp{\fontchoice}{helvetica}=0 %
	\usepackage[scaled]{helvet}
	
	\usepackage[T1]{fontenc}
\fi

\makeatother

\usepackage{ifthen}
\usepgfplotslibrary{fillbetween}

\definecolor{my_pink}{rgb}{1,0.07,0.65}
\definecolor{my_g}{rgb}{0.09,0.71,0.14}

\begin{document}

\title{
On Irradiance Distributions for Weakly Turbulent FSO Links: Log-Normal vs. Gamma-Gamma %On the Validity of the Log-Normal and Gamma-Gamma Distribution for FSO Links Under Weak Turbulence Conditions
}

% \author{Author name(s)}
% \address{Author affiliation and full address}
% \email{e-mail address}
%%Uncomment the following line to override copyright year from the default current year.
%\copyrightyear{2024}
\vspace{-3.5ex}
\author{Carmen Álvarez Roa\authormark{*}, Yunus Can Gültekin, Vincent van Vliet, Menno van den Hout, %Eduward Tangdiongga,
Chigo Okonkwo, and Alex Alvarado}

\address{Department of Electrical Engineering, Eindhoven University of Technology, Eindhoven, the Netherlands}

\email{\authormark{*}c.alvarez.roa@tue.nl} %% email address is required
\vspace{-3ex}
\begin{abstract}
Weak turbulence is commonly modeled using the log-normal distribution. Our experimental results show that this distribution fails to capture irradiance fluctuations in this regime. The Gamma-Gamma model is shown to be more accurate.
%The log-normal model is commonly assumed to be valid under weak turbulence. Our experimental results show that it fails to capture the measured irradiance fluctuations, whereas the Gamma-Gamma model remains accurate across all turbulence regimes.
%We experimentally analyze atmospheric turbulence statistics on a 4.6 km FSO link. Unlike previous shorter-distance works, our dataset spans Rytov variances $0.07$–$4$, showing Gamma-Gamma (GG) consistently outperforms log-normal (LN), even under weak turbulence conditions.
\end{abstract}
\vspace{-0.5ex}
\section{Introduction}
\vspace{-1ex}
Free-space optical (FSO) communications are considered a promising solution for establishing high-capacity point-to-point wireless links thanks to its large available unlicensed bandwidth. However, the performance of FSO systems is strongly impaired by atmospheric turbulence, which induces random irradiance fluctuations that can severely degrade the link reliability.%~\cite{Farid2007}.

The strength of turbulence is commonly characterized by the Rytov variance ($\sigma_\text{R}^2$).
Depending on its value, different statistical models have been proposed to describe irradiance fluctuations \cite[Fig.~7.4]{And05}. 
Under weak turbulence conditions ($\sigma_\text{R}^2<1$), the log-normal (LN) distribution is traditionally employed~\cite{Phillips:81}.  The Gamma-Gamma (GG) distribution, in contrast, is commonly used for strong turbulence ($\sigma_\text{R}^2>1$)~\cite{AlH01}.
In this work, we focus on the weak turbulence regime, where the scintillation index (SI)---a standard metric in FSO communications---provides a good estimate of the Rytov variance, i.e., $\sigma_R^2\approx \text{SI}$~\cite[Sec.~8]{And05}. Henceforth, we only consider SI in this paper.
 
The LN model is known to have limitations, particularly in capturing the tails of the probability density function (PDF) of the irradiance fluctuations. These tails are crucial for estimating the fade probability, i.e., the fraction of time the irradiance falls below a prescribed threshold~\cite[Sec.~9.9]{And05}.
Nevertheless, the LN model continues to be widely used in the weak turbulence regime, as evidenced by several studies~\cite{Farid2007, Yang_LN,Paper_using_LN, ESMAIL2021, Nazari2016}. Moreover, previous studies conducted on the same experimental link considered in this paper have also adopted the LN model specifically to obtain the Rytov variance~\cite{VanVlietOFC, VanVlietECOC}. 
By contrast, the GG distribution has been shown to provide good agreement with experimental measurements under strong turbulence~\cite{And05, AlH01}, but it is rarely applied in the weak turbulence regime. 
Previous experimental and numerical studies~\cite{ESMAIL2021, Churnside:89, HranilovicMarkov, And05, AlH01} have focused primarily on very weak turbulence regimes ($\text{SI} \leq 0.13$), leaving the \textit{intermediate weak turbulence regime} ($0.13 < \text{SI} \leq 1$) largely unexplored. Consequently, the accuracy of both LN and GG models in describing weak turbulence within this range of SIs has not been thoroughly verified experimentally.

This paper presents an extensive experimental study of turbulence-induced irradiance fluctuations over a $4.6$ km FSO link deployed above a dense urban area. This distance is considerably longer than those typically reported in previous studies (100 m–2.4 km)~\cite{ESMAIL2021,Nazari2016,Churnside:89,HranilovicMarkov}. %, allowing us to investigate the turbulence statistics over extended propagation paths.
%Our dataset spans a wide range of turbulence conditions. The observed variation in SI from $0.08$ to $4$ reflects both the accumulated turbulence effects along the link and the temporal variability of the atmosphere.
In this paper, we use experimental data within the \textit{intermediate weak turbulence regime} ($0.13 < \text{SI} \leq 1$)
to assess the validity of the LN and GG distributions.
To the best of our knowledge, this work provides the first experimental evidence from a deployed FSO system indicating that the LN model does not accurately approximate the PDF of the irradiance in this range. Our results also show that the GG model consistently provides a more accurate fit.

\vspace{-1.5ex}
\section{Experimental Setup and Signal Analysis}
\vspace{-1ex}
\begin{wrapfigure}[12]{r}{0.55\textwidth}
\vspace{-3.5ex}
\centering
\hspace{-2ex}
\scalebox{0.8}{\input{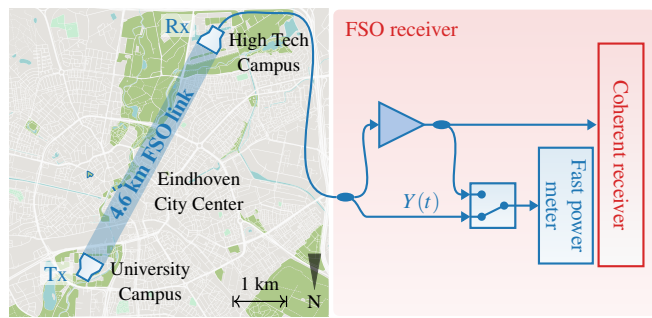}}
        \vspace{-4ex}
    \caption{Experimental setup for an FSO link connecting Eindhoven University of Technology and the High Tech Campus over 4.6 km.}
        \label{FigLink}
\end{wrapfigure}
A permanent FSO link has been established in Eindhoven, the Netherlands, connecting the Eindhoven University of Technology with the High Tech Campus. This link enables data transmission over a distance of $4.6$~km. The optical link is deployed above a dense urban area, directly crossing over the city center, as shown in Fig.~\ref{FigLink}~\cite{VanVlietOFC}. The optical terminals from the development partner Aircision are equipped with automated beam tracking and alignment systems tip/tilt correction, continuously optimizing the free-space-to-fiber coupling. As a result, pointing errors are largely compensated and their contribution to the observed signal fluctuations is minimized, allowing the analysis to focus on turbulence-induced fading.
A detailed description of the experimental setup can be found in \cite{VanVlietOFC, VanVlietECOC}. 

As shown in~Fig.~\ref{FigLink}, the received optical power was continuously monitored with a high-speed power meter operating at 10 kSa/s. Using this setup, a 30-hour measurement campaign was carried out on 14–15 October 2024.  
To characterize the temporal fluctuations of the channel, the recorded data was analyzed in consecutive, nonoverlapping, one-minute windows. Let $Y_k[n]$ be the $n$-th sample of the received optical power $Y(t)$ (see Fig.~\ref{FigLink}) within the $k$-th one-minute window. The SI %—a standard metric in FSO communications that quantifies the strength of atmospheric turbulence—
for the $k$th window is computed as
\vspace{-1.5ex}
\begin{equation}
    \text{SI}_k = \frac{\mathbb{E}\big\{Y_k[n]^2\big\} - \mathbb{E}\big\{Y_k[n]\big\}^2}{\mathbb{E}\big\{Y_k[n]\big\}^2},
\vspace{-1.5ex}
\label{eq:SI}
\end{equation}
%\vspace{-1ex}
where $\mathbb{E}\{\cdot\}$ indicates the expectation taken over the samples collected within the $k$-th window.
%The definition in \eqref{eq:SI} shows that SI is the normalized variance of the received signal. 
%Under weak turbulence conditions, the SI also provides a direct estimate of the Rytov variance, i.e., $\sigma_\text{R}^2 \approx \text{SI}$~\cite[Sec.~8]{And05}.

Following the approach of~\cite[Sec.~V]{Churnside:89}, we sort the values of $\text{SI}_k$ to group one-minute windows with similar levels of turbulence. This procedure enables the estimation of representative PDFs under different turbulence conditions.
To compensate for slow variations in the received optical power (e.g., caused by atmospheric attenuation), the signal within each one-minute window is normalized with respect to its mean. We therefore define the normalized samples of the received optical power $\overline{Y}_k[n]$ and their mean $\overline{\mu}_k$ as
\vspace{-0.5ex}
\begin{equation}
    \overline{Y}_k[n] \triangleq \frac{Y_k[n]}{\mathbb{E}\big\{Y_k[n]\big\}},
    \quad
    \overline{\mu}_k \triangleq \mathbb{E}\{10 \log (\overline{Y}_k[n])\}.
    \vspace{-1.5ex}
    \label{eq:NormalizePower}
\end{equation}
\vspace{-0ex}

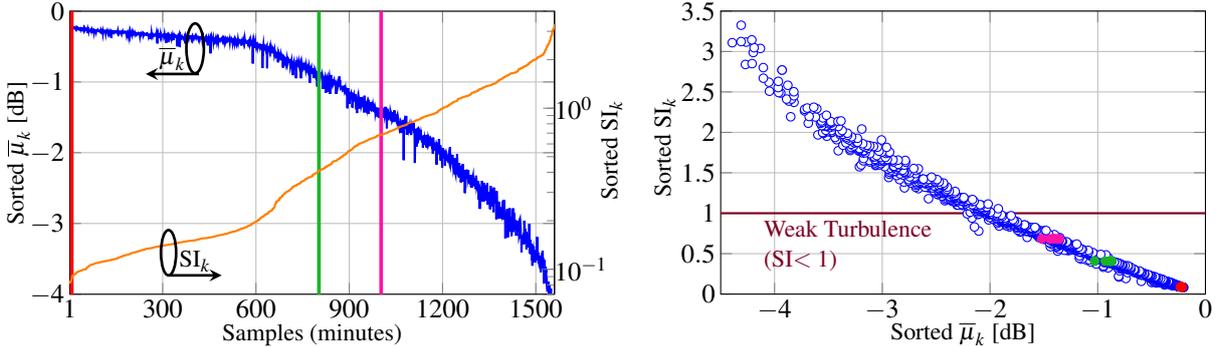
\begin{figure}[!h]
    % ----------- (bottom) Two plots side by side -----------
    \begin{subfigure}[t]{0.5\textwidth}
        \centering
    \pgfplotsset{compat=1.17}
\usetikzlibrary{arrows,decorations.markings}
\usetikzlibrary{calc,arrows}

\begin{tikzpicture}[scale=1]

\begin{axis}[ 
    width=1\textwidth,
    height=2.1in,   %alto de la figura
    xmin=1, xmax=1560,
    ymin=-4, ymax=0,  %similar a eje X
    label style={font=\small},
    xlabel={Samples (minutes)}, %Nombre eje X
    ylabel shift=-0ex,
    xlabel shift=0.5ex,
    ylabel={Sorted $\overline{\mu}_k$ [dB]}, %Nombre eje X
    xtick={1,300,600,...,1600},
     ytick={-4,-3,...,0},
     yticklabel style={xshift=-2pt} ,
    xticklabel style={yshift=-2pt},
     xticklabels={1,300,600,...,1500},
    grid=major% axis on top
]

\addplot [color=blue,solid,line width=0.8pt, mark options={solid},unbounded coords=jump,
    filter discard warning=false] file{./txtData/NormalizedPower_sorted.txt};\label{PowerDataSorted}
\draw[solid, thick, red] (axis cs:5,-4) -- (axis cs:5,0);
\draw[solid, thick, red] (axis cs:10,-4) -- (axis cs:10,0);

%\fill[my_g, solid, thick] (axis cs:800,-4) rectangle (axis cs:805,0);
\draw[solid, thick, my_g] (axis cs:800,-4) -- (axis cs:800,0);
\draw[solid, thick, my_g] (axis cs:805,-4) -- (axis cs:805,0);

\draw[solid, thick, my_pink] (axis cs:1000,-4) -- (axis cs:1000,0);
\draw[solid, thick, my_pink] (axis cs:1005,-4) -- (axis cs:1005,0);
%\addplot [color=blue,solid,line width=0.8pt, mark options={solid},unbounded coords=jump,filter discard warning=false] file{./txtData/ReceivedPower_sorted_average.txt};
    
\end{axis} %Finalizamos el diseño de la gráfica en sí

 \node [coordinate](input) {};
 
% \node[ellipse, color=black, line width=1pt, draw,minimum width = 0.6cm, minimum height = 0.05cm, rotate=90] (e) at (1.65,3.1) {};

%  \draw[-stealth,solid, black,line width=1pt,rounded corners]($(input.south)+(1.7,2.8)$) -- ($(input.south)+(1,2.8)$);
 
% \node[ellipse, color=black, line width=1pt, draw,minimum width = 0.6cm, minimum height = 0.05cm, rotate=90] (e) at (12.8,1.1) {};
%  \draw[-stealth,solid, black,line width=1pt,rounded corners]($(input.south)+(12.8,0.8)$) -- ($(input.south)+(13.5,0.8)$);

\begin{semilogyaxis}[
        width=1\textwidth,
        height=2.1in,   %alto de la figura 
        label style={font=\small},
        ylabel={Sorted $\text{SI}_k$},
        ymin=7e-2,ymax=4,
        xmin=1, xmax=1560,
        ylabel shift=-1ex,
        %domain=34:59,
        axis y line*=right,
        axis x line=none,
        %xtick distance =2
    %    ytick={5e-2,10e-1, 10e0 ,5,10e1},     % mejor lista de ticks
        xtick={1,200,400,...,1600}, 
         y label style={at={(axis description cs:1.14,0.51)}, anchor=south}, 
        ] 
   \addplot [color=orange,solid,line width=0.8pt, mark options={solid}, unbounded coords=jump,
    filter discard warning=false] file{./txtData/AverageSI_sorted.txt};\label{SIDataSorted}

\end{semilogyaxis}

%  \node [draw,fill=white,anchor= south west,font=\scriptsize] at (0.05,0.05) {\shortstack[l]{ 
% %\ref{QPSK} \hspace{0.01cm} QPSK\\
%  \ref{PowerData} $\Delta=0.20$~dB\\
%  \ref{SIData} $\Delta=0.19$~dB}};
  
\node [coordinate](input) {};

\node[ellipse, color=black, line width=1pt, draw,minimum width = 0.66cm, minimum height = 0.23cm, rotate=90] (e) at (1.65,3.26) {};

 \draw[-stealth,solid, black,line width=1pt,rounded corners]($(input.south)+(1.7,2.92)$) -- ($(input.south)+(1,2.92)$)node[midway, above, yshift=0.05cm, font=\small]{$\overline{\mu}_k$};
 
\node[ellipse, color=black, line width=1pt, draw,minimum width = 0.6cm, minimum height = 0.2cm, rotate=90] (e) at (1.3,0.55) {};

 \draw[-stealth,solid, black,line width=1pt,rounded corners]($(input.south)+(1.3,0.25)$) -- ($(input.south)+(2,0.25)$)node[midway, above, yshift=0.05cm,xshift=0.cm, font=\small]{$\text{SI}_k$};

%---------SIZE WIDTH 3.5----- (1.8, 4.1cm )
% \node at ($(input.north) + (1.8, 4.1cm )$) [scale=1, font=\scriptsize] {$\overline{\text{SNR}}_\text{e}\approx14.6$dB};

\end{tikzpicture}
        % \caption{With normalized power}
        % \label{FigSortedDataA}
    \end{subfigure}
    \hfill
    \hspace{1ex}
    \begin{subfigure}[t]{0.5\textwidth}
        \centering
            \usetikzlibrary{arrows,decorations.markings}
    \usetikzlibrary{calc,arrows}    
    
    \definecolor{my_pink}{rgb}{1,0.07,0.65}
\definecolor{my_g}{rgb}{0.09,0.71,0.14}
\definecolor{FEC}{rgb}{0.5, 0.0, 0.13}
    %\hspace{-0.7cm}
        \pgfplotsset{set layers} % �� activa las capas estándar de PGFPlot
    \begin{tikzpicture}[scale=1]    
    
   \begin{axis}[ 
       width=1\textwidth,
       height=2.1in,   %alto de la figura
        xmin=-4.5, xmax=0,
        ymin=0, 
        ymax=3.5,  
        label style={font=\small},
         %ylabel shift=-3ex,
        % xlabel shift=-1ex,
        ylabel={Sorted $\text{SI}_k$},
        xlabel={Sorted $\overline{\mu}_k$ [dB]}, %Nombre eje X
       ytick={0, 0.5, 1,...,3.5}, 
       grid=major,% axis on top
        y label style={at={(axis description cs:0.1,0.52)}, anchor=south}, 
        yticklabel style={xshift=-2pt} ,
        xticklabel style={yshift=-2pt},
        x label style={at={(axis description cs:0.5,0.04)}, anchor=north},
         z buffer=sort % <- importante para ordenar por capas
    ]

  \addplot[only marks, color=blue, mark=*, mark options={scale=0.8, draw=blue, fill=white},on layer=axis background] file {./txtData/CorrelationData_NormalizedPower.txt};

 \addplot[only marks, color=red, mark=*, mark options={scale=0.8, draw=red, fill=red},on layer=axis foreground] file {./txtData/Correlation/5_10minSamples_Correlation.txt};
 \addplot[only marks, color=my_g, mark=*, mark options={scale=0.8, draw=my_g, fill=my_g},on layer=axis foreground] file {./txtData/Correlation/800_805minSamples_Correlation.txt};
 \addplot[only marks, color=my_pink, mark=*, mark options={scale=0.8, draw=my_pink, fill=my_pink},on layer=axis foreground] file {./txtData/Correlation/1000_1005minSamples_Correlation.txt};
           %, filter discard warning=false,
    %x filter/.expression={mod(\coordindex,2)==0 ? x : nan}
 \draw[solid, thick, FEC, align=left] (axis cs:-5,1) -- (axis cs:0,1)node[below left, xshift=-3.6cm,yshift=-0.1cm,font=\small] {Weak Turbulence\\(SI$<1$)};;
    \end{axis} 

    \end{tikzpicture}
        % \caption{With non-normalized power}
        % \label{FigSortedDataB}
    \end{subfigure}

    \vspace{-4ex}
    % ----------- (global caption) -----------
    \caption{Sorted $\text{SI}_k$ and $\overline{\mu}_k$ (left), computed over one-minute windows and $\text{SI}_k$-$\overline{\mu}_k$ scatter plot (right).}
    \label{FigSI_Power}
\end{figure}

Fig.~\ref{FigSI_Power} presents the relationship between the sorted values of $\text{SI}_k$ from \eqref{eq:SI} and their mean normalized power $\overline{\mu}_k$ from \eqref{eq:NormalizePower}. Fig.~\ref{FigSI_Power} (left) shows the evolution of the sorted data. % sorting the data into one-minute windows of increasing $\text{SI}_k$, 
%allowing the identification of periods with similar turbulence strength. 
A clear opposite trend is observed: higher SI values are associated with lower mean normalized power levels. The use of normalized power in~\eqref{eq:NormalizePower} effectively removes the influence of long-term drifts or absolute power variations, resulting in a clearer and more physically meaningful correlation, which is not readily apparent in \cite[Fig.~2]{VanVlietOFC}. Fig.~\ref{FigSI_Power} (right) presents the corresponding scatter plot (blue circles). Fig.~\ref{FigSI_Power} (right) shows a clear correlation between $\text{SI}_k$ and $\overline{\mu}_k$.
As shown in Fig.~\ref{FigSI_Power}, the dataset corresponding to the 30-hour measurement period considered in this work spans both weak and strong turbulence regimes. In the following section, we focus exclusively on the weak turbulence regime ($\text{SI}<1$).
%In the following section, the normalized empirical PDF is compared with the analytical PDFs of the LN and GG models.
%-------------------
\vspace{-1ex}

\vspace{-5.5ex}
\section{Numerical Results}
\vspace{-1ex}
The LN and GG distributions are parametric distributions. We estimate these parameters ($\sigma^2$ for LN and $(\alpha,\beta)$ for GG) using maximum likelihood estimation~\cite[$\mbox{Sec.~IV-C}$]{HranilovicMarkov}.
The sorted data is grouped into windows with similar SIs, following the method proposed in~\cite{Churnside:89}. In our case, five consecutive one-minute windows are combined to form each grouped dataset. This five-minute grouping provides enough samples to accurately estimate the tails of the PDFs while keeping the parameters of the PDFs approximately constant during each dataset.

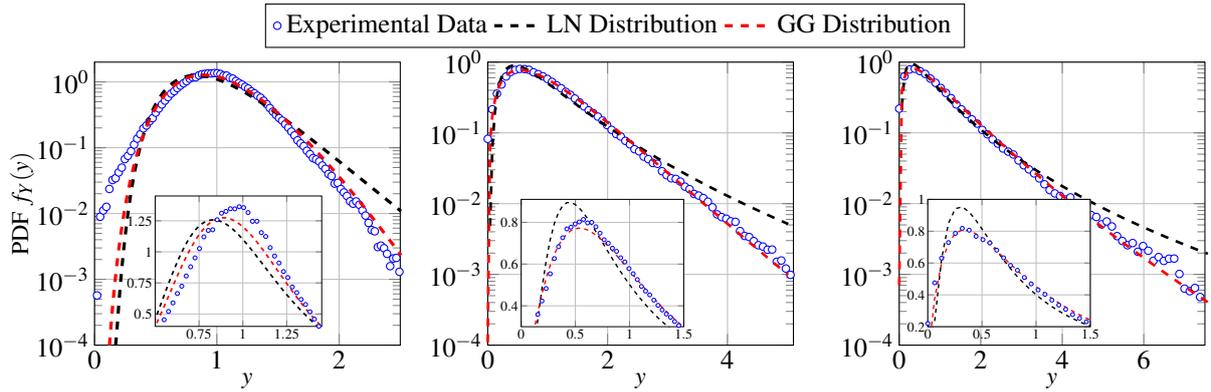
\begin{figure}[!t]
    \centering
    \parbox{\textwidth}{\centering
        \pgfplotslegendfromname{GlobalLegend}
    }
    % --- Top row (log scale) ---
    \begin{subfigure}[t]{0.32\linewidth}
        \centering
%% SIGMA_R^2=0.101

    \usetikzlibrary{arrows,decorations.markings}
    \usetikzlibrary{calc,arrows}    
    
    \definecolor{my_pink}{rgb}{1,0.07,0.65}
\definecolor{my_g}{rgb}{0.09,0.71,0.14}
\definecolor{FEC}{rgb}{0.5, 0.0, 0.13}
    %\hspace{-0.7cm}

    \pgfplotsset{set layers} % �� activa las capas estándar de PGFPlot
    
    \begin{tikzpicture}[scale=1]    
    
   \begin{semilogyaxis}[ 
       width=1.1\textwidth,
       height=2.1in,   %alto de la figura
         label style={font=\small},  
        xmin=0, xmax=2.5,
        ymin=1e-4, ymax=2,  
        xlabel={$y$}, %Nombre eje X
        ylabel shift=-2ex,
       ytick={0.0001,0.001,0.01,0.1,1},
       yticklabel style={xshift=-2pt} ,
        xticklabel style={yshift=-2pt},
        ylabel={PDF $f_Y(y)$},
             grid=major,% axis on top
            ylabel style={
           at={(-0.01,0.5)},     
           anchor=south, 
           yshift=-3ex
       },
       legend style = {at={(0.5,0.04)}, anchor=south},
        legend columns=-1,
        legend entries={Experimental Data, LN Distribution, GG Distribution},
        legend to name={GlobalLegend},
        xlabel style={
            at={(0.5,0)},      % 0.5 = centro horizontal del eje, 0 = borde inferior del eje
            anchor=north,      % ancla la etiqueta hacia arriba
            yshift=1ex         % mueve la etiqueta hacia arriba (positivo)
        },
         z buffer=sort % <- importante para ordenar por capas
    ]

  \addplot[only marks, color=blue, mark=*, mark options={scale=0.7, draw=blue, fill=white}, on layer=axis background, filter discard warning=false,
    x filter/.expression={mod(\coordindex,2)==0 ? x : nan}] file {./NewData_PDF/Window5to10min/Empirical_pdf.txt};\label{empirical}

\addplot [color=black,dashed,line width=1.2pt,mark options={solid},on layer=axis foreground] file{./NewData_PDF/Window5to10min/LN_MLE_pdf.txt};\label{LN_MLE}
           
\addplot [color=red,dashed,line width=1.2pt,on layer=axis foreground,mark options={solid}] file{./NewData_PDF/Window5to10min/GG_MLE_pdf.txt};\label{GG}

    \end{semilogyaxis}

% \node [draw,fill=white,anchor= south west,font=\scriptsize] at (0.7,0.1) {\shortstack[l]{ 
% \hspace{0.17cm} \ref{empirical} \hspace{0.2cm}   Experimental Data\\
% \ref{LN_MLE} LN Distribution\\
% \ref{GG}  GG Distribution }};

    \node [coordinate](input) {};

    \end{tikzpicture}   
    \end{subfigure}
    \hfill
    \begin{subfigure}[t]{0.32\linewidth}
        \centering
            \usetikzlibrary{arrows,decorations.markings}
    \usetikzlibrary{calc,arrows}    
    
    \definecolor{my_pink}{rgb}{1,0.07,0.65}
\definecolor{my_g}{rgb}{0.09,0.71,0.14}
\definecolor{FEC}{rgb}{0.5, 0.0, 0.13}
    %\hspace{-0.7cm}
    \pgfplotsset{set layers} % �� activa las capas estándar de PGFPlot
    \begin{tikzpicture}[scale=1]    
    
   \begin{semilogyaxis}[ 
       width=1.1\textwidth,
         label style={font=\small},  
       height=2.1in,   %alto de la figura
        xmin=0, xmax=5.1,
        ymin=1e-4, 
        ymax=1,  
        xlabel={$y$}, %Nombre eje X
        ylabel shift=-2ex,
        yticklabel style={xshift=-2pt} ,
        xticklabel style={yshift=-2pt},
        ylabel={},
        ytick={0.0001,0.001,0.01,0.1,1},
         grid=major,% axis on top
            ylabel style={
           at={(0,0.5)},      % coloca el label en x=0 (borde del eje), y=0.5 (centro vertical)
           anchor=south,      % controla la orientación
           yshift=-3ex
       },
       xlabel style={
            at={(0.5,0)},      % 0.5 = centro horizontal del eje, 0 = borde inferior del eje
            anchor=north,      % ancla la etiqueta hacia arriba
            yshift=1ex         % mueve la etiqueta hacia arriba (positivo)
        },
         z buffer=sort % <- importante para ordenar por capas
    ]
 
\addplot[only marks, color=blue, mark=*, mark options={solid,scale=0.7,fill=white}, on layer=axis background,filter discard warning=false,x filter/.expression={mod(\coordindex,2)==0 ? x : nan}] file {./NewData_PDF/Window800to805min/Empirical_pdf.txt};

\addplot [color=black,dashed,line width=1pt, on layer=axis foreground,mark options={solid}] file{./NewData_PDF/Window800to805min/LN_MLE_pdf.txt};
           
\addplot [color=red,dashed,line width=1pt,on layer=axis foreground,mark options={solid}] file{./NewData_PDF/Window800to805min/GG_MLE_pdf.txt};
       
\end{semilogyaxis}

% \node [draw,fill=white,anchor= south west,font=\scriptsize] at (0.3,0.1) {\shortstack[l]{ 
% \hspace{0.17cm} \ref{empirical} \hspace{0.2cm}   Experimental Data\\
% \ref{LN_MLE} LN Distribution\\
% \ref{GG}  GG Distribution }};

    \end{tikzpicture}
    \end{subfigure}
    \hfill
    \begin{subfigure}[t]{0.32\linewidth}
        \centering
            \usetikzlibrary{arrows,decorations.markings}
    \usetikzlibrary{calc,arrows}    
    
    \definecolor{my_pink}{rgb}{1,0.07,0.65}
\definecolor{my_g}{rgb}{0.09,0.71,0.14}
\definecolor{FEC}{rgb}{0.5, 0.0, 0.13}

    \pgfplotsset{set layers} % �� activa las capas estándar de PGFPlot
    \begin{tikzpicture}[scale=1]    
    
   \begin{semilogyaxis}[ 
       width=1.1\textwidth,
       height=2.1in,   %alto de la figura
         label style={font=\small},  
        xmin=0, xmax=7.5,
        ymin=1e-4, 
        ymax=1,  
        xlabel={$y$}, %Nombre eje X
        ylabel shift=-2ex,
        ylabel={},
        yticklabel style={xshift=-2pt} ,
        xticklabel style={yshift=-2pt},
        ytick={0.0001,0.001,0.01,0.1,1},
         grid=major,% axis on top
            ylabel style={
           at={(0,0.5)},      % coloca el label en x=0 (borde del eje), y=0.5 (centro vertical)
           anchor=south,      % controla la orientación
           yshift=-3ex
       },
       xlabel style={
            at={(0.5,0)},      % 0.5 = centro horizontal del eje, 0 = borde inferior del eje
            anchor=north,      % ancla la etiqueta hacia arriba
            yshift=1ex         % mueve la etiqueta hacia arriba (positivo)
        },
         z buffer=sort % <- importante para ordenar por capas
    ]

   \addplot[only marks, color=blue, mark=*, mark options={solid,scale=0.7,fill=white}, on layer=axis background,filter discard warning=false,
    x filter/.expression={mod(\coordindex,2)==0 ? x : nan}] file {./NewData_PDF/Window1000to1005min/Empirical_pdf.txt};

\addplot [color=black,dashed,line width=1pt,on layer=axis foreground,mark options={solid}] file{./NewData_PDF/Window1000to1005min/LN_MLE_pdf.txt};

\addplot [color=red,dashed,line width=1pt,on layer=axis foreground,mark options={solid}] file{./NewData_PDF/Window1000to1005min/GG_MLE_pdf.txt};

    \end{semilogyaxis}

% \node [draw,fill=white,anchor= south west,font=\scriptsize] at (0.2,0.1) {\shortstack[l]{ 
% \hspace{0.17cm} \ref{empirical} \hspace{0.2cm}   Experimental Data\\
% \ref{LN_MLE} LN Distribution\\
% \ref{GG}  GG Distribution }};

    \node [coordinate](input) {};

    \end{tikzpicture}
    \end{subfigure}

    \vspace{-2.6cm} % space between rows
    % --- Bottom row (linear scale) ---
    \hspace{-1.5ex}
    \begin{subfigure}[b]{0.325\linewidth}
        \centering
        \resizebox{0.55\linewidth}{!}{    \usetikzlibrary{arrows,decorations.markings}
    \usetikzlibrary{calc,arrows}    
    
    \definecolor{my_pink}{rgb}{1,0.07,0.65}
\definecolor{my_g}{rgb}{0.09,0.71,0.14}
\definecolor{FEC}{rgb}{0.5, 0.0, 0.13}
    %\hspace{-0.7cm}
    \begin{tikzpicture}[scale=1]    
    
   \begin{axis}[ 
       width=1.1\textwidth,
       height=1.9in,   %alto de la figura
        xmin=0.5, xmax=1.45,
        ymin=0.4, 
        ymax=1.45,  
        yticklabel style={xshift=-2pt} ,
        xticklabel style={yshift=-2pt},
       % xlabel={Normalized channel gain}, %Nombre eje X
       % ylabel={PDF},
        ytick={0.5,0.75,1,1.25},
        xtick={0.75,1,1.25},
          grid=major,
          axis background/.style={fill=white}  % �� 
    ]

  \addplot[only marks, color=blue, mark=*, mark options={scale=0.7, draw=blue, fill=white}, filter discard warning=false,
    x filter/.expression={mod(\coordindex,2)==0 ? x : nan}] file {./NewData_PDF/Window5to10min/Empirical_pdf.txt};

\addplot [color=black,dashed,line width=1.2pt,mark options={solid}] file{./NewData_PDF/Window5to10min/LN_MLE_pdf.txt};
           
\addplot [color=red,dashed,line width=1.2pt,mark options={solid}] file{./NewData_PDF/Window5to10min/GG_MLE_pdf.txt};
   
    \end{axis}

% \node [draw,fill=white,anchor= south west,font=\scriptsize] at (1.2,2.45) {\shortstack[l]{ 
% \hspace{0.17cm} \ref{empirical} \hspace{0.2cm}   Experimental Data\\
% \ref{LN_MLE} LN Distribution\\
% \ref{GG}  GG Distribution }};

    \node [coordinate](input) {};

    \end{tikzpicture}}
    \end{subfigure}
    %\hfill
    \hspace{-3ex}
    \begin{subfigure}[b]{0.325\linewidth}
        \centering
         \resizebox{0.55\linewidth}{!}{
            \usetikzlibrary{arrows,decorations.markings}
    \usetikzlibrary{calc,arrows}    
    
    \definecolor{my_pink}{rgb}{1,0.07,0.65}
\definecolor{my_g}{rgb}{0.09,0.71,0.14}
\definecolor{FEC}{rgb}{0.5, 0.0, 0.13}
    %\hspace{-0.7cm}
    \begin{tikzpicture}[scale=1]    
    
   \begin{axis}[ 
       width=1.1\textwidth,
       height=1.9in,   %alto de la figura
        xmin=0, xmax=1.5,
        ymin=0.3, 
        ymax=0.91,  
        ylabel shift=-2ex,
        yticklabel style={xshift=-2pt} ,
        xticklabel style={yshift=-2pt},
         grid=major,% axis on top
            ylabel style={
           at={(-0.05,0.5)},   
           anchor=south,     
           yshift=-4ex
       },
       xlabel style={
            at={(0.5,0)},      % 0.5 = centro horizontal del eje, 0 = borde inferior del eje
            anchor=north,      % ancla la etiqueta hacia arriba
            yshift=1ex         % mueve la etiqueta hacia arriba (positivo)
        },
        axis background/.style={fill=white}  
    ]
 
   \addplot[only marks, color=blue, mark=*, mark options={solid,scale=0.7,fill=white}, %filter discard warning=false,x filter/.expression={mod(\coordindex,2)==0 ? x : nan}
   ] file {./NewData_PDF/Window800to805min/Empirical_pdf.txt};
                  
\addplot [color=black,dashed,line width=1pt,mark options={solid}] file{./NewData_PDF/Window800to805min/LN_MLE_pdf.txt};

\addplot [color=red,dashed,line width=1pt,mark options={solid}] file{./NewData_PDF/Window800to805min/GG_MLE_pdf.txt};

    \end{axis} 

% \node [draw,fill=white,anchor= south west,font=\scriptsize] at (1.2,2.45) {\shortstack[l]{ 
% \hspace{0.17cm} \ref{empirical} \hspace{0.2cm}   Experimental Data\\
% \ref{LN_MLE} LN Distribution\\
% \ref{GG}  GG Distribution }};
  
    \end{tikzpicture}}
    \end{subfigure}
   % \hfill
    \hspace{0ex}
    \begin{subfigure}[b]{0.325\linewidth}
        \centering
         \resizebox{0.55\linewidth}{!}{
            \usetikzlibrary{arrows,decorations.markings}
    \usetikzlibrary{calc,arrows}    
    
    \definecolor{my_pink}{rgb}{1,0.07,0.65}
\definecolor{my_g}{rgb}{0.09,0.71,0.14}
\definecolor{FEC}{rgb}{0.5, 0.0, 0.13}
    %\hspace{-0.7cm}
    \begin{tikzpicture}[scale=1]    
    
   \begin{axis}[ 
       width=1.1\textwidth,
       height=1.9in,   %alto de la figura
        xmin=0, xmax=1.5,
        ymin=0.2, 
        ymax=1,  
        yticklabel style={xshift=-2pt} ,
        xticklabel style={yshift=-2pt},
        ylabel shift=-2ex,
         grid=major,% axis on top
            ylabel style={
           at={(-0.05,0.5)},      % coloca el label en x=0 (borde del eje), y=0.5 (centro vertical)
           anchor=south,      % controla la orientación
           yshift=-4ex
       },
       xlabel style={
            at={(0.5,0)},      % 0.5 = centro horizontal del eje, 0 = borde inferior del eje
            anchor=north,      % ancla la etiqueta hacia arriba
            yshift=1ex         % mueve la etiqueta hacia arriba (positivo)
        },
        axis background/.style={fill=white}
        ]
 
   \addplot[only marks, color=blue, mark=*, mark options={solid,scale=0.7,fill=white}] file {./NewData_PDF/Window1000to1005min/Empirical_pdf.txt};

\addplot [color=black,dashed,line width=1pt,mark options={solid}] file{./NewData_PDF/Window1000to1005min/LN_MLE_pdf.txt};

\addplot [color=red,dashed,line width=1pt,mark options={solid}] file{./NewData_PDF/Window1000to1005min/GG_MLE_pdf.txt};
      
    \end{axis}

% \node [draw,fill=white,anchor= south west,font=\scriptsize] at (1.2,2.45) {\shortstack[l]{ 
% \hspace{0.17cm} \ref{empirical} \hspace{0.2cm}   Experimental Data\\
% \ref{LN_MLE} LN Distribution\\
% \ref{GG}  GG Distribution }};

    \node [coordinate](input) {};

    \end{tikzpicture}}
    \end{subfigure}
    \vspace{2.5ex}
    \caption{PDF of the the normalized received optical power $f_Y(y)$ under weak turbulence for experimental data. Results for the LN and GG distributions are also shown. 
    Columns correspond to $\mbox{$\text{SI} = 0.101$}$ (left),  $\text{SI} = 0.486$ (middle), and $\text{SI} = 0.817$ (right).}
    %BEFORE $\sigma_R^2 = 0.098$,Figures/Window1000to1005
    %(center) $\sigma_R^2 = 0.266$, 
    %and (right) $\sigma_R^2 = 0.672$
    \vspace{-4ex}
    \label{FigResults}
\end{figure}

To illustrate representative turbulence conditions, three datasets were selected, corresponding to $\text{SI}=0.101$, $\text{SI}=0.486$, and $\text{SI}=0.817$. These datasets are indicated in Fig.~\ref{FigSI_Power} (left) by the vertical red, green, and pink lines, resp., and are highlighted in Fig.~\ref{FigSI_Power} (right) as colored circles using the same color scheme.
Fig.~\ref{FigResults} shows the corresponding PDFs obtained from these datasets. In each plot, the circles represent the normalized empirical PDF, while the dashed black and red curves correspond to the LN and GG distributions, resp.. The three subplots correspond to $\text{SI}=0.101$ (left), $\text{SI}=0.486$ (middle), and $\text{SI}=0.817$ (right).
The horizontal axis represents the normalized received optical power within each five-minute data window.
A logarithmic scale is used on the vertical axis to better visualize the tails of the PDFs, while the inset with linear scaling provides a clearer view of the peak of the PDFs.

The results for $\text{SI}=0.101$ (Fig.~\ref{FigResults}, left) provide an experimental validation of the conclusions reported in \mbox{\cite{And05,AlH01}} for $\text{SI}\approx0.1$, which relied exclusively on simulation results. Consistent with these works, our measurements show that both the LN and GG models underestimate the left tail of the PDF. Fig.~\ref{FigResults} (left) also shows that the GG model provides a noticeably better fit to the right tail of the PDF.
Fig.~\ref{FigResults} (middle) and Fig.~\ref{FigResults} (right) present the results for $\text{SI} = 0.486$ and $\text{SI} = 0.817$, respectively. %, illustrating the behavior of the models at \textit{intermediate weak} turbulence levels.
These results show that the LN distribution underestimates both the peak and the right tail of the empirical PDFs, whereas the GG distribution closely matches the experimental data. Overall, across all three turbulence conditions, the GG distribution provides a significantly better fit to the empirical data than the LN model, particularly for the tails of the PDFs.

%Our experimental results therefore demonstrate the limitations of the LN model in the weak turbulence regime, where it is traditionally applied, and highlight the need for more accurate models. The GG distribution, commonly used for strong turbulence, provides an accurate fit in the weak regime.

%such as the GG distribution, for realistic FSO communications system analysis.

%For~$\sigma_R^2=0.098$ the data is expected to be LN. However, from Fig.~\ref{FigResults}(left, top), we see that the LN PDF underestimates the probability density in the tails of the PDF \ref{ANDREEEW}. Moreover, the linear plot in Fig. \ref{FigResults}(left, bottom) shows that the log-normal PDF underestimates the peak probability density. 
\vspace{-1ex}

\vspace{-0.5ex}
\section{Conclusions}
\vspace{-1ex}

We investigated the validity of the LN distribution for modeling weak atmospheric turbulence using measurements from a $4.6$ km experimental FSO link crossing directly over the city center of Eindhoven. In contrast to previous studies, our dataset spans a much wider range of SI, extending into the \textit{intermediate weak turbulence regime} ($0.13 < \text{SI} \leq 1$) that has remained largely unexplored.
Our analysis showed that, within this regime, the empirical PDFs exhibit clear deviations from the LN model. 
Specifically, the LN distribution overestimates the peak of the PDF and underestimates the distribution tails, which could lead to inaccurate predictions of deep fades, and, consequently, of outage probability. While similar trends have been observed for other SI ranges in numerical simulations, no prior experimental or numerical evidence exists for the SI range considered here. Our results provide direct experimental evidence that the LN model is inadequate for accurately describing weak turbulence conditions, where it is traditionally assumed to work well.
Moreover, our results show that the GG model, often used only for strong turbulence, provides a more accurate fit across weak turbulence conditions, outperforming the LN model throughout the SI range considered. 

\vspace{1.5ex}
\noindent\footnotesize{\textbf{Acknowledgements:} This publication is part of the project BIT-FREE with file number 20348 of the research programme Open Technology Programme and of the project AI-SUSAT with grant ID OAARE97526 of the research programme AiNed XS Europe, which are (partly) financed by the Dutch Research Council (NWO). This research has received funding from the Project Optical Wireless Superhighways: Free photons (FREE) under Grant P19-13, the PhotonDelta National Growth Fund Programme on Photonics, and the European Innovation Council Transition project CombTools under Grant G.A. 101136978. The authors thank Aircision B.V., particularly Nourdin Kaai, Roland Blok, and Andreas Kotilis, for their support in setting up the High Tech Campus location of the Reid Photonloop FSO testbed. The authors also thank Prof. Eduward Tangdiongga for his key contributions to the setup of the testbed.% under the grant https://www.doi.org/10.61686/????.
}

\end{document}